\documentclass[12pt]{article}

\usepackage{bbm} 
\usepackage{hyperref} 
\usepackage{enumitem} 
\usepackage{algorithm, algorithmicx, algpseudocode} 
\usepackage{graphicx} 
\usepackage{amsmath} 
\usepackage{amsfonts} 
\usepackage{mathtools} 
\usepackage{booktabs} 
\usepackage{natbib} 
\DeclareMathOperator{\E}{E}
\DeclareMathOperator{\Var}{Var}
\DeclareMathOperator{\supp}{supp}

\begin{document}

\title{Distilling Importance Sampling for Likelihood Free Inference}

\author{Dennis Prangle\footnote{dennis.prangle@bristol.ac.uk}\\
School of Mathematics, University of Bristol, UK\\
Cecilia Viscardi\\
Department of Statistics, Computer Science, Applications\\
University of Florence, Italy}

\date{}

\maketitle

\bigskip

\begin{abstract}
Likelihood-free inference involves inferring parameter values given observed data and a simulator model.
The simulator is computer code which takes parameters,
performs stochastic calculations, and outputs simulated data.
In this work, we view the simulator as a function whose inputs are (1) the parameters and
(2) a vector of pseudo-random draws.
We attempt to infer all these inputs conditional on the observations.
This is challenging as the resulting posterior can be high dimensional and involve strong dependence.

We approximate the posterior using normalizing flows, a flexible parametric family of densities.
Training data is generated by likelihood-free importance sampling with a large bandwidth value $\epsilon$,
which makes the target similar to the prior.
The training data is ``distilled'' by using it to train an updated normalizing flow.
The process is iterated, using the updated flow as the importance sampling proposal,
and slowly reducing $\epsilon$ so the target becomes closer to the posterior.
Unlike most other likelihood-free methods, we avoid the need to reduce data to low dimensional
summary statistics, and hence can achieve more accurate results. We illustrate our
method in two challenging examples, on queuing and epidemiology.

\end{abstract}

\section{Introduction} \label{sec:intro}

Many statistical models are specified by \emph{simulators}, which can be used to generate data under the model given values of parameters.
Probability calculations are often not possible for such models.
In particular, it can be infeasible to evaluate the \emph{likelihood function} $L(\theta)$:
the probability (or density) of the observed data under parameters $\theta$.
This makes it difficult to use standard methods of statistical inference
such as maximum likelihood and many Monte Carlo methods for Bayesian inference.

\emph{Likelihood-free inference} (LFI)
-- also known as \emph{simulation-based inference} --
describes methods to perform inference using simulators without evaluating the likelihood function.
One popular class of LFI methods is \emph{approximate Bayesian computation} (ABC) \citep{Marin:2012}.
Here one runs the simulator under many $\theta$ values,
and each time calculates a distance between the simulated data $y$ and the observed data $y_0$.
The distances are used to produce a sample (or weighted sample) of $\theta$s from an approximation to the posterior
e.g.~by selecting the $\theta$s with the smallest distances.

A drawback of ABC is that is suffers from a \emph{curse of dimensionality}:
the cost of producing an accurate posterior approximation rises rapidly with $\dim(y)$.
An intuitive explanation is that, even under the best $\theta$ choices,
close matches of $y$ and $y_0$ are rare unless $\dim(y)$ is low.
Therefore it is common to use dimension reduction,
replacing $y$ with a vector of low dimensional \emph{summary statistics} $s(y)$
when calculating distances.
However, using summary statistics typically results in a loss of posterior accuracy.

An alternative class of LFI methods use \emph{conditional density estimation} \citep{Grazian:2019}.
These estimate a density based on simulated pairs of parameters and data.
Then one can condition on the observed data to approximate its posterior distribution.
These methods avoid the ABC curse of dimensionality,
but typically still require dimension reduction of the data to summary statistics.
So the associated loss of information remains a problem.

We aim to improve the efficiency of LFI algorithms by \emph{parameter augmentation}.
The idea is to infer not only the model parameters $\theta$,
but also all the random variables the simulator samples during its operation.
Effectively this learns to control the simulator to produce output similar to the observations.
Now it is no longer the case that outputting close matches will always be rare.
In principle this can avoid the ABC curse of dimensionality without the need for summary statistics.

However, inference under parameter augmentation is challenging.
The posterior is high dimensional and may have strong dependencies
e.g.~lying on a lower dimensional manifold.
This makes it hard to design effective proposals for Monte Carlo methods.
We use approximate versions of the posterior, which correspond to inference under noisy observation of the data,
with a \emph{bandwidth} parameter $\epsilon$ controlling the variance of the noise.
The resulting distribution is the prior for $\epsilon \to \infty$,
and the posterior for $\epsilon \to 0$.
Thus large $\epsilon$ produces a less challenging inference problem.

We approximate these distributions using \emph{normalizing flows} \citep{Papamakarios:2019},
a family of flexible and computationally tractable distributions.
They transform a random vector from a simple random distribution to a random vector from a complex distribution,
by applying a sequence of learnable bijections.
Normalizing flows can be trained using batches of data sampled from a target distribution.

We propose a method which alternates two steps.
The first is importance sampling, using the current normalizing flow as the proposal.
The target distribution is an approximate posterior with large $\epsilon$,
chosen so that importance sampling produces reasonably accurate results.
The second step is to use the resulting weighted sampled to train the normalizing flow further.
Following \citet{Li:2017}, we refer to this as \emph{distilling} the importance sampling results.
By iteratively distilling importance sampling results, we can target increasingly accurate posterior approximations i.e.~reduce $\epsilon$.

We demonstrate our algorithm on two challenging examples.
In a queuing example, we obtain approximate inference results which are much more accurate than two ABC baselines:
with and without summary statistics.
In an epidemiology example, we are able to target the exact posterior.


In the remainder of the paper, Section \ref{sec:background} presents background material.
Section \ref{sec:augmentation} discusses LFI parameter augmentation.
Sections \ref{sec:objective} and \ref{sec:alg} describe our method.
Section \ref{sec:sin} illustrates it on a simple two dimensional inference task.
Sections \ref{sec:MG1} and \ref{sec:SI} present our main examples.
Section \ref{sec:conclusion} concludes with a discussion, including limitations and opportunities for extensions.
Further details are available in the supplement, including a discussion of recommended tuning choices (Section \ref{sec:tuning summary}).
Code for the examples can be found at \url{https://github.com/dennisprangle/DistillingImportanceSampling},
written using PyTorch \citep{Paszke:2019}.
All examples were run on a 16-core desktop PC.

\subsection{Related Work and Novelty} \label{sec:related}

Several recent papers \citep{Muller:2019, Arbel:2021, Duan:2019, Naesseth:2020}
train a normalizing flow to use as an importance sampling proposal.
Variations include training a Gaussian mixture \citep{Jerfel:2021}
or a distribution transformed using polynomials \citep{Cotter:2019},
rather than the usual neural network approaches in normalizing flows literature (reviewed in Section \ref{sec:flows}).
The novelty of our work is an application to likelihood-free inference.
To do so, we sequentially target a sequence of approximate distributions, unlike the prior work listed above.

Parameter augmentation methods for likelihood-free inference have been used previously.
\cite{Graham:2017} propose a method using constrained Hamiltonian Monte Carlo (HMC).
This alternates HMC steps with projection steps to move the MCMC state back onto a target manifold.
A limitation of this method is that it requires a differentiable simulator.
Optimisation Monte Carlo (OMC) \citep{Meeds:2015} generates random seeds to use within the simulator,
then uses optimisation to find the corresponding $\theta$ values which minimise the distance of the resulting $y$ to $y_0$,
and computes appropriate weights.
Robust OMC \citep{Ikonomov:2020} produces multiple $\theta$s given a random seed.
Rare event ABC \citep{Prangle:2018} is a MCMC algorithm which, whenever a new $\theta$ is proposed, uses rare event methods to estimate the probability of the simulator producing a close match.

Let $x$ denote the internal random variables used by the simulator
(defined more precisely in Section \ref{sec:augmentation}.)
Then OMC methods essentially sample $x$ from its prior,
which could be inefficient if the posterior for $x$ is concentrated compared to the prior.
Rare event ABC infers $p(x | \theta, y)$ for each proposed $\theta$, which becomes expensive for long MCMC chains.
In contrast to these two methods, our approach aims to infer $p(\theta, x | y)$,
which we argue is more efficient.

Joint inference of $\theta$ and $x$
can also sometimes be performed by sophisticated MCMC algorithms specialised to particular applications.
For instance \citet{Shestopaloff:2014} give an MCMC algorithm for the queuing example we consider in Section \ref{sec:MG1}.
Our goal in this paper is to provide a more generic approach.

More broadly, our approach has connections to several inference methods.
Concentrating on its importance sampling component, it is closely related to
\emph{adaptive importance sampling} \citep{Cornuet:2012}.
Concentrating on training an approximate density,
it is related to the \emph{cross-entropy method} \citep{Rubinstein:1999},
an \emph{estimation of distribution} algorithm \citep{Larranaga:2002},
and \emph{reweighted wake-sleep} \citep{Bornschein:2014}.

\section{Background} \label{sec:background}


\subsection{Likelihood-Free Inference} \label{sec:framework}

Suppose we observe data $y_0$, assumed to be the output of a probability model with density $p(y|\theta)$ under some parameters $\theta$,
and we have access to a prior density $\pi(\theta)$.
Bayesian inference aims to find the corresponding posterior density
$p(\theta|y_0) \propto \pi(\theta) p(y_0|\theta)$.
When the \emph{likelihood function} $p(y|\theta)$ can be evaluated,
many approaches to posterior inference are available.

However, many models are specified using a \emph{simulator}, typically some computer code.
The simulator's inputs are parameters $\theta$, and the output is data $y$.
This effectively defines a distribution\footnotemark{} for $y$ conditional on $\theta$.
However, probability calculations under this distribution are typically not feasible,
including evaluation of the likelihood function.
So standard likelihood-based methods of inference are not possible.
\footnotetext{If the simulator is deterministic the distribution is a point mass.
However it is more common to consider stochastic simulators with LFI.}
\emph{Likelihood-free inference} (LFI) methods attempt to perform approximate inference
in this setting without directly evaluating the likelihood function.

One popular LFI approach is approximate Bayesian computation (ABC).
The simplest ABC algorithm, \emph{rejection ABC}, samples $(\theta, y)$ pairs from the prior and simulator,
and accepts those where a distance between $y$ and $y_0$
(e.g.~the Euclidean distance $||y-y_0||$ if data is a vector of fixed length)
is below some threshold $\epsilon$.
The accepted $\theta$s form a sample from an approximation to the posterior.
More efficient ABC algorithms exist,
for instance using more sophisticated methods of sampling $\theta$ values,
or using importance sampling ideas to allow weighted samples.

Despite these improvements, close matches of $y$ and $y_0$ are rare unless $\dim(y)$ is low.
As a consequence, ABC suffers from a \emph{curse of dimensionality}:
in the asymptotic case $\epsilon \to 0$, the cost of ABC rapidly increases with $\dim(y)$.
So ABC typically uses dimension reduction.
That is, acceptance now occurs when $||s(y) - s(y_0)||$ is small,
for some function $s$ mapping data to a vector of \emph{summary statistics}.
The loss of information entailed by dimension reduction reduces inference accuracy
(as sufficient statistics are rarely available).

See \cite{Prangle:2018} (end of Section 2.1) for a brief review of the ABC curse of dimensionality,
and \cite{Marin:2012} for a review of other relevant aspects of ABC.
The supplement (Section \ref{app:ABCalg}) gives more details of a specific ABC algorithm
we use as a baseline in one of our examples.

As mentioned in Section \ref{sec:intro},
an alternative approach to LFI is through \emph{conditional density estimation} methods (CDE), as reviewed by \cite{Grazian:2019}.
CDE methods estimate a density, often a normalizing flow, using simulated $(\theta, y)$ pairs.
The density estimated could be the joint $\pi(\theta,y)$ or a conditional: $p(\theta | y)$ or $p(y | \theta)$.
One can then condition on the observed data to approximate its posterior distribution.
The method for conditioning depends on which density was estimated.
CDE methods avoid the ABC curse of dimensionality,
but typically still require dimension reduction of the data to summary statistics.
So the resulting loss of information remains a problem.

\subsection{Importance Sampling} \label{sec:IS}

Let $p(\xi) = \tilde{p}(\xi)/Z$ be a \emph{target density} where only $\tilde{p}(\xi)$ can be evaluated,
and the value of the normalizing constant $Z = \int \tilde{p}(\xi) d\xi$ is unknown.
Importance sampling (IS) is a Monte Carlo method to estimate expectations of the form
$I = \E_{\xi \sim p}[h(\xi)]$,
for some function $h$.
Here we review relevant aspects.
For full details see e.g.~\citet{Robert:2013} and \citet{Rubinstein:2016}.

IS requires a \emph{proposal density} $\lambda(\xi)$ which can easily be sampled from, and must satisfy
\begin{equation} \label{eq:supp}
\supp(p) \subseteq \supp(\lambda),
\end{equation}
where $\supp$ denotes support.
Then
\begin{equation} \label{eq:IS}
I  = \E_{\xi \sim \lambda} \left[ \frac{p(\xi)}{\lambda(\xi)} h(\xi) \right].
\end{equation}
So an unbiased Monte Carlo estimate of $I$ is
\begin{equation} \label{eq:ISunnorm}
\hat{I}_1 = \frac{1}{NZ} \sum_{i=1}^N w_i h(\xi^{(i)}),
\end{equation}
where $\xi^{(1)}, \xi^{(2)}, \ldots, \xi^{(N)}$ are independent samples from $\lambda$, and $w_i = \tilde{p}(\xi^{(i)})/\lambda(\xi^{(i)})$ is an \emph{importance weight}.

Typically $Z$ is estimated as $\tfrac{1}{N} \sum_{i=1}^N w_i$ giving
\begin{equation} \label{eq:ISnorm}
\hat{I}_2 = \sum_{i=1}^N w_i h(\xi^{(i)}) \bigg/ \sum_{i=1}^N w_i,
\end{equation}
a biased, but consistent, estimate of $I$.
Equivalently
\[
\hat{I}_2 = \sum_{i=1}^N s_i h(\xi^{(i)}),
\]
for \emph{normalized} importance weights $s_i = w_i / \sum_{i=1}^N w_i$.

A drawback of IS is that it can produce estimates with large, or infinite, variance if $\lambda$ is a poor approximation to $p$.
Hence diagnostics for the quality of the results are useful.
A popular diagnostic is \emph{effective sample size} (ESS),
\begin{equation} \label{eq:ESS}
N_{\text{ESS}} = \left(\sum_{i=1}^N w_i\right)^2 \bigg/ \ \sum_{i=1}^N w_i^2.
\end{equation}
Under various assumptions, $\Var(\hat{I}_2)$ roughly equals the variance of an idealized Monte Carlo estimate based on $N_{\text{ESS}}$ independent samples from $p(\xi)$ \citep{Liu:1996}.
\cite{Elvira:2022}, amongst others,
note that ESS can be an unreliable diagnostic in practice
and research is needed to propose better alternatives.

\subsection{Normalizing Flows} \label{sec:flows}

A \emph{normalizing flow} represents a random vector $\xi$ with a complicated distribution
as an invertible transformation of a random vector $z$ with a simple \emph{base distribution},
typically $\mathcal{N}(0,\mathrm{I})$.

Recent research has developed flexible learnable families of normalizing flows.
See \citet{Papamakarios:2019} for a review.
We focus on autoregressive neural spline flows \citep{Durkan:2019},
which we will refer to as ``spline flows'' as a shorthand.
See Section \ref{sec:q_choice} for comments on exploratory work with an alternative choice.

\paragraph{Spline Flows Description}

An autoregressive transformation transforms input vector $u$ to output vector $v$
using $v_i = \tau(u_i, h(u_{<i}))$
(where $u_{<i}$ refers to the $u_j$ values with $j<i$).
Here $\tau$ is a monotonic and invertible transformation of its first argument.
The particular transformation used depends on the second argument.
So $v_i$ is a transformation of $u_i$,
and the transformation used depends only on $u_{<i}$.
Therefore the overall transformation has a triangular Jacobian matrix,
facilitating quick density calculations.
Furthermore, a masked autoregressive neural network \citep{Germain:2015, Nash:2019}
is typically used for $h$.
This allows $h(u_{<i})$ to be computed in parallel for all $i$ values.

\cite{Durkan:2019} propose using a spline transformation for $\tau$.
This transformation is a piecewise function based on partitioning a bounding box $[-B,B]$ into several bins.
The type of function used is chosen so it can be fully defined by:
the location of knots (bin boundaries);
the function values at the knots;
and (in some cases) derivatives at the knots.
All these details should be the output of $h(u_{<i})$.
We use piecewise rational quadratic transformations, following \cite{Durkan:2019}.
Outside the bounding box the function is defined to be the identity.

\paragraph{Spline Flows Properties}

Some relevant properties of spline flows are as follows.
See \cite{Papamakarios:2019} for full discussion of the points made here.
\begin{itemize}
\item
\textbf{Universality.}
An autoregressive flow can represent any distribution meeting some simple conditions,
if any $\tau$ and $h$ functions can be used.
In practice this means the expressiveness of spline flows is only limited by the
complexity of the splines and neural networks used.
\item
\textbf{Sampling.}
Samples of $\xi$ can rapidly be drawn from $q(\xi; \phi)$.
\item
\textbf{Gradient calculation.}
It is reasonably fast to compute $\nabla \log q(\xi; \phi)$.
(Throughout the paper $\nabla$ represents the gradient operator with respect to $\phi$.)
\end{itemize}

Universality means that spline flows can approximate many distributions well.
Sampling and gradient calculation are required in our algorithm.
Alternative types of flow can allow faster gradient calculation, but are often less expressive.

We note that Gaussian mixture models are an alternative to flows.
However their cost of sampling and density evaluation grows rapidly with $\dim \xi$,
as it involve a $O([\dim \xi]^3)$ matrix inversion cost.

\section{Parameter Augmentation} \label{sec:augmentation}

Here we outline an approach to likelihood-free inference which our algorithm will use.
The idea is to infer not just the parameters $\theta$,
but also all the internal stochastic behaviour of the simulator.
To do so, we introduce a random variable $x$.
This represents all outputs of an underlying random number generator used by the simulator.
We suppose all the sampled random variables required during simulation can be expressed as transformations of $\theta$ and $x$.
So the simulator is now a deterministic function $y(\xi)$,
where $\xi$ represents the \emph{augmented parameters} $(\theta, x)$.

Throughout this paper we work with simulators where $x$ can be expressed as a fixed length random vector $x \sim \mathcal{N}(0,\mathrm{I})$.
We denote its density as $\pi(x)$.
Future work on more complex simulators could consider alternative choices of $x$.
Firstly, $x$ could be allowed to be dependent on $\theta$.
Secondly, $x$ could be an infinite sequence of independent $\mathcal{N}(0,1)$ random variables.
This is helpful if the number of random draws required by the simulator is unknown in advance.
For instance the simulator could perform a loop for a number of iterations which depends on $\theta$ in a complex fashion.
Another possible choice is to allow $x$ to be dependent on $\theta$.


\subsection{Advantages and Challenges}

An advantage of parameter augmentation is avoiding the ABC curse of dimensionality.
We first give an intuitive explanation.
Sampling $\xi$ from the posterior $p(\xi | y_0)$, or a close approximation,
can be viewed as \emph{controlling the simulator} through $x$ so that $y(\xi)$ is a close match to $y_0$.
This avoids the problem of close matches being rare when only $\theta$ is controlled,
as discussed in Section \ref{sec:framework}.
Secondly, we give a more direct explanation.
Suppose we can produce a good approximation $q(\xi)$ of the augmented posterior\footnotemark{} $p(\xi | y)$,
which is suitable as an importance sampling proposal.
This allows straightforward inference using standard importance sampling.
\footnotetext{Or of an approximation $p_\epsilon(\xi)$ as defined below in \eqref{eq:LFItarget}.}

A difficulty of parameter augmentation is that the posterior for $\xi$
often has a challenging form.
A first challenge is that $\xi$ usually has high dimension.
\cite{Mackay:2003} argues that importance sampling is not suitable for high dimensional targets
due to the difficulty of producing suitable proposals.
Recent advances in approximating distributions has made progress on this challenge.
For instance \cite{Muller:2019} demonstrates that normalizing flows can learn good importance sampling proposals
for difficult target distributions, including high dimensional cases.
A theoretical justification for this is the universality property discussed in Section \ref{sec:flows}.
This is the motivation for using flows in this paper.

A second challenge is that parameter augmention may result in extremely strong posterior dependence.
Indeed the posterior is often a singular distribution whose mass lies on a lower dimensional manifold \citep{Graham:2017}.
Monte Carlo inference for such posteriors is challenging due to the difficulty of producing proposals exactly on an unknown manifold.
We make progress by performing inference on a sequence of increasingly accurate approximate posteriors,
which are all non-singular.

\subsection{Approximate Posteriors} \label{sec:LFItempering}

We define approximate posterior densities for $\xi$, controlled by a \emph{bandwidth} value $\epsilon > 0$,
as $p_\epsilon(\xi) = \tilde{p}_\epsilon(\xi) / Z_\epsilon$ where
\begin{align} 
\tilde{p}_\epsilon(\xi) &= 
\pi(\xi) \exp \left[ -\tfrac{1}{2 \epsilon^2} ||y(\xi) - y_0||^2 \right] \label{eq:LFItarget} \\
Z_\epsilon &= \int \tilde{p}_\epsilon(\xi) d\xi. \nonumber
\end{align}
Here $\pi(\xi) = \pi(\theta) \pi(x)$
and $||y(\xi) - y_0||$ is the Euclidean distance between the simulated and observed data.
Densities of the form \eqref{eq:LFItarget} are often used in ABC
and correspond to the posterior for data observed with independent $\mathcal{N}(0,\epsilon^2)$ additive errors \citep{Wilkinson:2013}.

Other similar approximate posteriors can be defined.
For instance, the most common ABC approximation 
takes $\tilde{p}_\epsilon(\xi) = \pi(\xi) \mathbbm{1}[ ||y(\xi) - y_0|| \leq \epsilon ]$,
where $\mathbbm{1}$ denotes an indicator function.
We prefer the approximation \eqref{eq:LFItarget} in this paper since it has the same support as $\pi(\xi)$.
This makes the occurrence of zero importance weights in our algorithm less likely (or impossible if $\pi(\xi)$ has full support).
Such zero weights can cause numerical problems to our algorithm.

It will be useful later to extend the unnormalised density \eqref{eq:LFItarget} to $\epsilon = 0$ by taking
\begin{equation}
\tilde{p}_0(\xi) = \pi(\xi) \mathbbm{1}[y(\xi) = y_0]. \label{eq:exactLFItarget}
\end{equation}
A valid density $p_0$ results when $Z_0 > 0$.
This is typically the case when the data $y$ is discrete, but not when it is continuous.
When $Z_0 = 0$, the distribution for $\epsilon \to 0$ is singular with respect to $\pi(\xi)$.

\section{Objective and Gradient} \label{sec:objective}

Our algorithm approximates $p_\epsilon$ using a family of densities $q(\xi; \phi)$, typically a normalizing flow.
This section introduces an objective function to judge how well $q$ approximates $p_\epsilon$.
It then discusses how to estimate the gradient of this objective with respect to $\phi$.
Section \ref{sec:alg} presents our algorithm,
which uses these gradients to update $\phi$ while also reducing $\epsilon$.

\subsection{Objective}

Given $p_\epsilon$, we aim to minimize the inclusive Kullback-Leibler (KL) divergence,
\begin{equation} \label{eq:incKL}
KL(p _\epsilon|| q) = \E_{\xi \sim p_\epsilon}  [\log p_\epsilon(\xi) - \log q(\xi; \phi)].
\end{equation}
This is equivalent to maximizing a scaled negative \emph{cross-entropy},
which is our objective,
\begin{equation*}
\mathcal{J}_\epsilon(\phi) = Z_\epsilon \E_{\xi \sim p_\epsilon} [ \log q(\xi; \phi) ].
\end{equation*}
(Scaling by $Z_\epsilon$ avoids this intractable constant appearing in our gradient estimates below.)

The inclusive KL divergence penalizes $\phi$ values which produce small $q(\xi; \phi)$ when $p_\epsilon(\xi)$ is large.
Hence the optimal $\phi$ tends to make $q(\xi; \phi)$ non-negligible where $p_\epsilon(\xi)$ is non-negligible, known as the \emph{zero-avoiding} property.
This is an intuitively attractive feature for importance sampling proposal distributions.
Indeed recent theoretical work shows that, under some conditions, the sample size required in importance sampling scales exponentially with the inclusive KL divergence \citep{Chatterjee:2018}.

Our work could be adapted to use the $\chi^2$ divergence \citep{Dieng:2017, Muller:2019},
which 
also has theoretical links to the sample size needed by IS \citep{Agapiou:2017}.

\subsection{Basic Gradient Estimate} \label{sec:gradients}

Assuming standard regularity conditions \citep[][Section 4.3.1]{Mohamed:2019},
the objective has gradient
\begin{equation} \label{eq:Jgradient}
\nabla \mathcal{J}_\epsilon(\phi) = Z_\epsilon \E_{\xi \sim p_\epsilon} [ \nabla \log q(\xi; \phi) ].
\end{equation}
Using \eqref{eq:IS}, an importance sampling form is
\begin{equation*}
\nabla \mathcal{J}_\epsilon(\phi) = \E_{\xi \sim \lambda} \left[ \frac{\tilde{p}_\epsilon(\xi)}{\lambda(\xi)} \nabla \log q(\xi; \phi) \right],
\end{equation*}
where $\lambda(\xi)$ is a proposal density satisfying \eqref{eq:supp}.
For now we assume $\lambda$ is not a function of $\phi$.
An unbiased Monte Carlo gradient estimate of \eqref{eq:Jgradient} is
\begin{equation} \label{eq:J grad est}
g_1 = \frac{1}{N} \sum_{i=1}^N w_i \nabla \log q(\xi^{(i)}; \phi),
\end{equation}
where $\xi^{(i)} \sim \lambda(\xi)$ are independent samples and
$w_i = \tilde{p}_\epsilon(\xi^{(i)}) / \lambda(\xi^{(i)})$
(importance sampling weights).
We calculate $\nabla \log q(\xi^{(i)}; \phi)$ by backpropagation (see e.g.~\citealp{Baydin:2018}).

\paragraph{Choice of Proposal Density}

In our main algorithm, defined below in Section \ref{sec:alg}, when gradient estimates are required
we have an existing estimate of $\phi$
(i.e.~at the start of the outer for loop in Algorithm \ref{alg:DIS} we have access to $\phi_{t-1}$.)
Therefore it is appealing to take $\lambda(\xi) = q(\xi; \phi)$.
Since we use choices of $q$ with full support, \eqref{eq:supp} is satisfied.

This $\lambda$ is a function of $\phi$, so the $\xi^{(i)}$ terms may have some dependence on $\phi$.
Interpreting $\nabla$ as full differentiation could cause terms involving $\nabla \xi^{(i)}$ to appear in $g_1$,
preventing it being an unbiased estimate of $\nabla \mathcal{J}_\epsilon(\phi)$.
To avoid this henceforth we take $\nabla$ as the gradient operator based on \emph{partial} differentiation with respect to $\phi$.
(Or equivalently we take $\lambda(\xi) = q(\xi; \bot \phi)$,
where $\bot$ is the ``stop-gradient'' operator in the notation of \citealp{Foerster:2018}.
This matches the implementation in code, as $\bot$ corresponds to the torch command \texttt{detach}.)

\subsection{Improved Gradient Estimates} \label{sec:improved grads}

Here we discuss reducing the variance and cost of $g_1$.

\paragraph{Truncating Weights}

To avoid high variance gradient estimates we apply \emph{truncated importance sampling} \citep{Ionides:2008}.
This truncates the weights at a maximum value $\omega$,
giving truncated importance weights $\tilde{w}_i = \min(w_i, \omega)$.
The resulting gradient estimate is
\[
g_2 = \frac{1}{N} \sum_{i=1}^N \tilde{w}_i \nabla \log q(\xi^{(i)}; \phi).
\]
This typically has lower variance than $g_1$, but has some bias.
See the supplement (Section \ref{app:truncation}) for more details and discussion,
including how we choose $\omega$ automatically,
and a heuristic argument why truncation bias is not problematic.

One can also think of $g_2$ as replacing 
$\tilde{p}_\epsilon(\xi^{(i)})$ with
$\min[\tilde{p}_\epsilon(\xi), \omega \lambda(\xi)]$.
Thus we attempt to optimize $q$ in a local neighbourhood of $\lambda$.

A more sophisticated alternative to truncation
is Pareto smoothing the largest importance weights \citep{Yao:2018}.
We do not use this as it is more expensive to implement than truncation,
but it would be interesting to investigate in future work.

\paragraph{Resampling}

Calculating $g_2$ requires evaluating $\nabla \log q(\xi^{(i)}; \phi)$ for $1 \leq i \leq N$.
Each of these has a computational cost, but often many receive small weights and so contribute little to $g_2$.

To reduce this cost we can discard many low weight samples, by using \emph{importance resampling} \citep{Smith:1992} as follows.
We sample $n \ll N$ times, with replacement, from the $\xi^{(i)}$s with probabilities $\tilde{s}_i = \tilde{w}_i/S$ where $S = \sum_{i=1}^N \tilde{w}_i$.
Denote the resulting samples as $\tilde{\xi}^{(j)}$.
Then an unbiased estimate of $g_2$ is
\begin{equation} \label{eq:J grad est3}
g_3 = \frac{S}{nN} \sum_{j=1}^n \nabla \log q(\tilde{\xi}^{(j)}; \phi).
\end{equation}


\section{Algorithm} \label{sec:alg}

Algorithm \ref{alg:DIS} gives our main inference algorithm,
and this section discusses various details of it.
The algorithm outputs an $\epsilon$ value and a density $q(\xi; \phi)$ trained to be an importance sampling proposal
for $p_\epsilon(\xi)$.
We recommend using $q$ in a final stage of importance sampling with a large sample size.
While the density $q$ is itself an approximation of $p_\epsilon$ (and therefore of the exact posterior),
in our experience it can have artefacts: see discussion at the end of Section \ref{sec:sin}.

\begin{algorithm}[htbp] \caption{Distilled importance sampling (DIS)} \label{alg:DIS}
\begin{algorithmic}[1]
\State Input: importance sampling size $N$, target ESS $M$, batch size $n$, number of batches $B$.
\State Initialize $\phi_0$ (followed by pretraining if necessary) and let $\epsilon_0 = \infty$.
\For{$t = 1,2,\ldots$}
\State Sample $(\xi_i)_{1 \leq i \leq N}$ from $q(\xi; \phi_{t-1})$.
\State Select a new value $\epsilon_t \leq \epsilon_{t-1}$
(see Section \ref{sec:updating eps} for details).
\State Calculate $w_i = \tilde{p}_\epsilon(\xi^{(i)}) / q(\xi^{(i)}; \phi_{t-1})$ weights
and truncate to $\tilde{w}_i$s (see the supplement, Section \ref{app:truncation}, for details).
\For{$j = 1,2,\ldots,B$}
\State Resample $(\tilde{\xi}^{(j)})_{1 \leq j \leq n}$ from $(\xi^{(i)})_{1 \leq i \leq N}$ using normalized $\tilde{w}_i$s as probabilities, with replacement.\!\! 
\State Calculate gradient estimate $g$ using \eqref{eq:J grad est3}.
\State Calculate $\phi_{t,j}$ from $\phi_{t,j-1}$
and $g$ using a step of an optimization algorithm such as Adam
(where $\phi_{t,0} = \phi_{t-1}$.)
\EndFor
\State Let $\phi_{t} = \phi_{t,B}$.
\If{$\epsilon = 0$ or runtime above prespecified limit}
\State \Return $q(\xi; \phi_t)$ (typically to be used as importance sampling proposal)
\EndIf
\EndFor
\end{algorithmic}
\end{algorithm}

\subsection{Algorithm Overview}

A summary of the algorithm follows.
Steps 4--6 perform importance sampling with target $p_\epsilon$.
Steps 7--11 use the output to train $\phi$, aiming for the resulting density to approximate the importance sampling target.
As the algorithm progresses, step 5 reduces $\epsilon$,
aiming for the importance sampling ESS to equal $M$ (a tuning choice).
The goal is to make the importance sampling target closer to the posterior,
slowly enough to avoid high variance gradient estimates.

To use the importance sampling output efficiently,
we sample from it (with replacement) several times to create training batches,
and use each batch for one update of $\phi$.
We fix training batch size to $n=100$,
and number of batches $B$ to $\lceil M/n \rceil$.
So approximately $M$ importance sampling outputs are used for training.
Limiting the number of outputs used aims to avoid overfitting from too much reuse of the same training data.

An update of $\phi$ is done by a step of an optimisation algorithm based on gradient estimates,
aiming to increase $\mathcal{J}_\epsilon$.
The gradient estimate is calculated from the current batch of training data,
and is calculated using \eqref{eq:J grad est3}.
We use the Adam optimisation algorithm \citep{Kingma:2014}, and found it worked well,
but many alternatives could also be used
(see the review of \citealp{Ruder:2016} for instance).

Steps 13--14 terminate the algorithm after a fixed runtime or once $\epsilon=0$ is reached
(when this is possible i.e.~for discrete data.)
Alternatively, the termination decision could be based on approximate inference diagnostics,
such as those of \cite{Yao:2018} or \cite{Huggins:2020}.

We investigate the remaining tuning choices empirically in Section \ref{sec:MG1}.
For now note that $N$ must be reasonably large since our method to update $\epsilon_t$ relies on making an accurate ESS estimate, as detailed in Section \ref{sec:updating eps}.
A further summary and discussion of tuning choices appears in the supplement (Section \ref{sec:tuning summary}).

\subsection{Pretraining} \label{sec:init}

Our initial target is the prior $\pi(\xi)$, since we take $\epsilon_0 = \infty$.
The initial $q$ should be similar to this.
Otherwise the first gradient estimates produced by importance sampling are likely to have high variance.
To achieve this we often make use of \emph{pretraining}.
We assume the prior can easily be sampled from.
Then pretraining iterates the following steps:
\begin{enumerate}
\item Sample $(\xi^{(i)})_{1 \leq i \leq n}$ from $\pi(\xi)$.
\item Update $\phi$ using gradient
$\frac{1}{n} \sum_{i=1}^n \nabla \log q(\xi^{(i)}; \phi)$.
\end{enumerate}
This maximizes the negative cross-entropy
$\E_{\xi \sim \pi} [ \log q(\xi; \phi) ]$.
We use $n=100$,
and terminate once $q(\xi; \phi)$ achieves a ESS of 75
when targeting the prior in importance sampling.

\subsection{Selecting $\epsilon_t$} \label{sec:updating eps}

We select $\epsilon_t$ using ESS, as in \citet{DelMoral:2012}.
Given $(\xi_i)_{1 \leq i \leq N}$ sampled from $q(\xi;\phi_{t-1})$,
the ESS value for target $p_\epsilon(\xi)$ is
\begin{align*}
N_{\text{ESS}}(\epsilon) &= \left[ \sum_{i=1}^N w(\xi_i, \epsilon) \right]^2
\bigg/ \sum_{i=1}^N w(\xi_i, \epsilon)^2 \\
\text{where} \ w(\xi, \epsilon) &= \tilde{p}_\epsilon(\xi) / q(\xi;\phi_{t-1}).
\end{align*}
(Also, to avoid numerical errors, we define the ESS to be zero if all weights are zero.)

In step 5 of Algorithm \ref{alg:DIS} we first check whether $N_{\text{ESS}}(\epsilon_{t-1}) < M$, the target ESS value.
If so we set $\epsilon_t = \epsilon_{t-1}$.
Otherwise we set $\epsilon_t$ to an estimate of the minimal $\epsilon$ such that $N_{\text{ESS}}(\epsilon) \geq M$,
computed by a bisection algorithm\footnotemark.
\footnotetext{
We must sometimes bisect intervals of the form $[a, \infty]$.
To do so we take the midpoint to be $a+100$.
}
We perform at least 50 iterations of bisection, and then terminate once
$N_{\text{ESS}}(\epsilon_t) \in [M-0.01, M+0.01]$.

\subsection{Reaching $\epsilon_t = 0$: Exact Inference} \label{sec:exact_info}

For continuous data,
the set of $\xi$ such that $y(\xi) = y_0$ typically has probability zero\footnotemark{}
under any choice of $q(\theta; \phi)$.
Hence DIS cannot reach $\epsilon=0$ since the unnormalised density $\tilde{p}_0$ from \eqref{eq:exactLFItarget}
will almost surely result in all weights being zero.
Instead, like ABC, DIS produces increasingly good posterior approximations as $\epsilon \to 0$.
\footnotetext{For instance because this set has Lebesgue measure zero and $q$ has Lebesgue dominating measure.}

However for discrete data,
it is plausible to reach $\epsilon_t=0$.
When this happens, the algorithm produces an approximation of the augmented posterior $p_0(\xi)$
which is a suitable proposal density for exact importance sampling.

\subsection{Choice of $q$} \label{sec:q_choice}

For all our examples we use an autoregressive neural spline flow for $q(\xi; \phi)$ with 5 bins for each variable, and bounding box $[-10,10]$.
The spline details are output by an autoregressive residual neural network \citep{Nash:2019}
with 3 blocks, 20 hidden features and ReLU activation.
More comments on the choice of $q$ are in Section \ref{sec:tuning summary} of the supplement.

\section{Illustration: Sinusoidal Simulator} \label{sec:sin}

As a simple illustration,
consider the simulator $y(\theta, x) = -\sin \theta + x$
with independent priors $\theta \sim U(-\pi,\pi)$, $x \sim \mathcal{N}(0,1)$.
As observations we take $y_0 = 0$.
Thus the exact posterior is supported on the line $x = \sin \theta$.

It's helpful to reparameterise $\theta$ to a distribution with unbounded support.
So we introduce $\vartheta \sim \mathcal{N}(0,1)$
and let $\theta = \pi [2 \Phi(\vartheta) - 1]$
where $\Phi$ is the $\mathcal{N}(0,1)$ cumulative distribution function.
We then infer $\xi=(\vartheta, x)$.

We use Algorithm \ref{alg:DIS} with $N=4000$ training samples and a target ESS of $M=2000$.
These values give a clear visual illustration: we investigate efficient tuning choices later.
We pretrain so that $q(\xi;  \phi_0)$ approximates the prior.

\begin{figure}[tb]
\begin{center}
\includegraphics[width=0.3\textwidth]{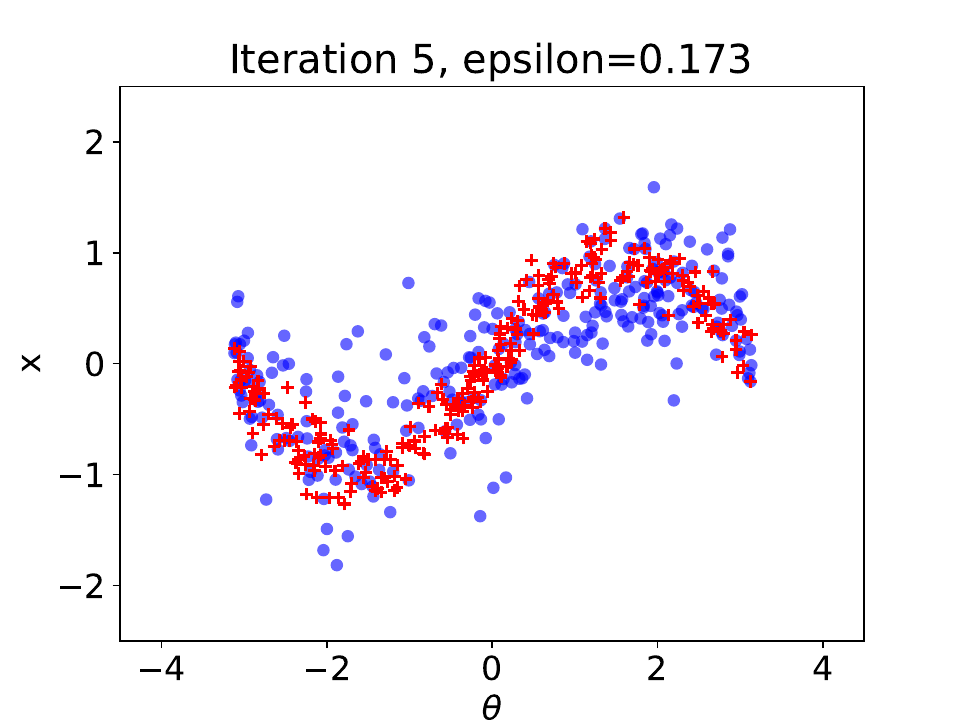}
\includegraphics[width=0.3\textwidth]{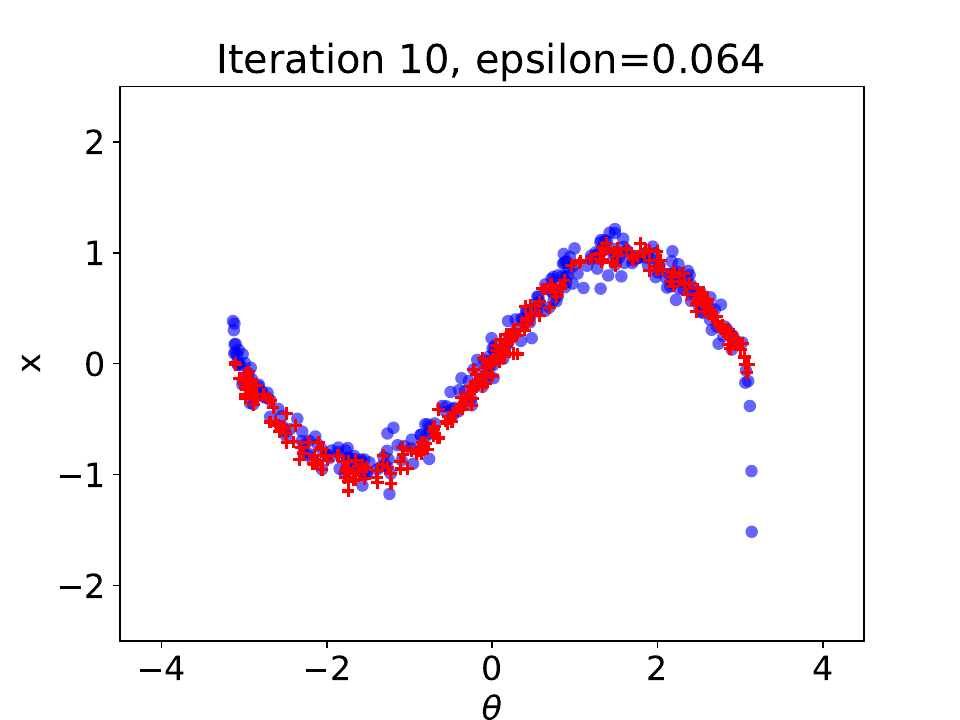}
\includegraphics[width=0.3\textwidth]{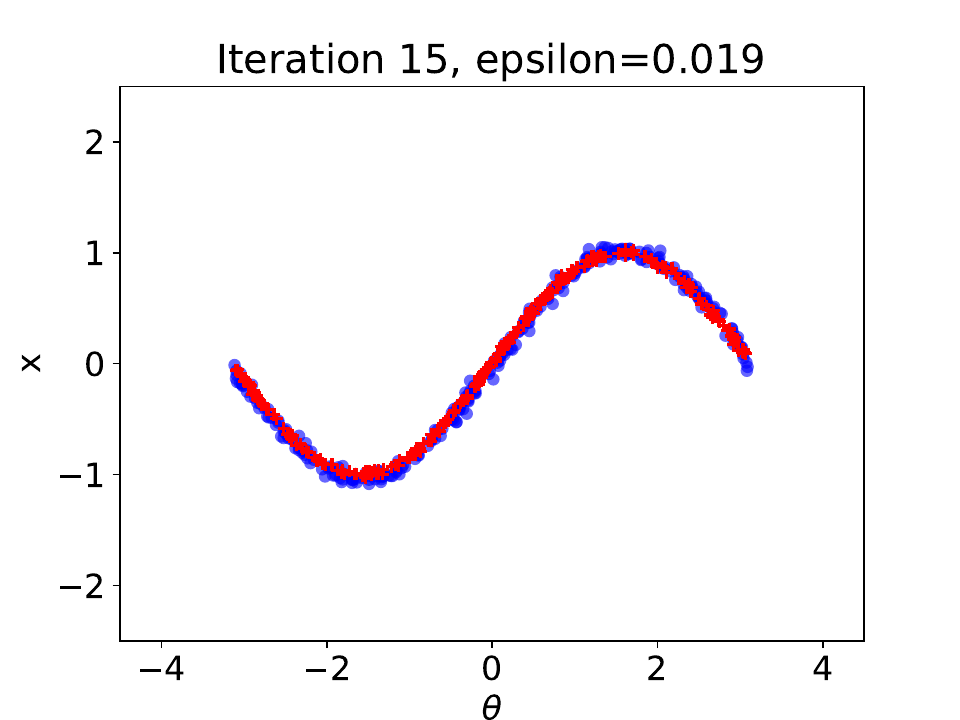}\\
\includegraphics[width=0.3\textwidth]{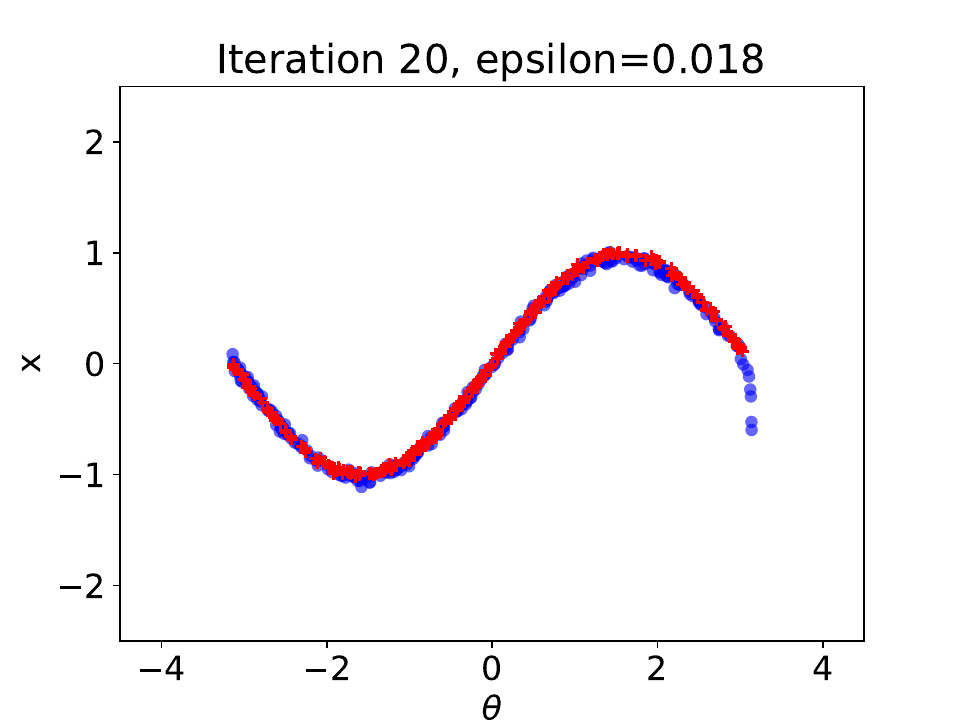}
\includegraphics[width=0.3\textwidth]{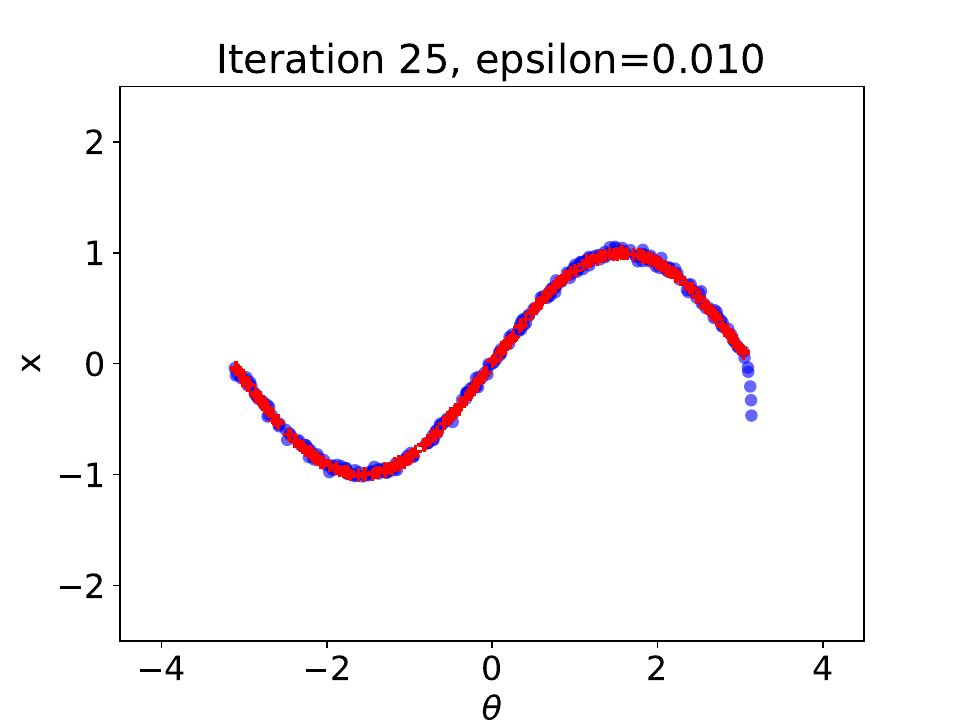}
\includegraphics[width=0.3\textwidth]{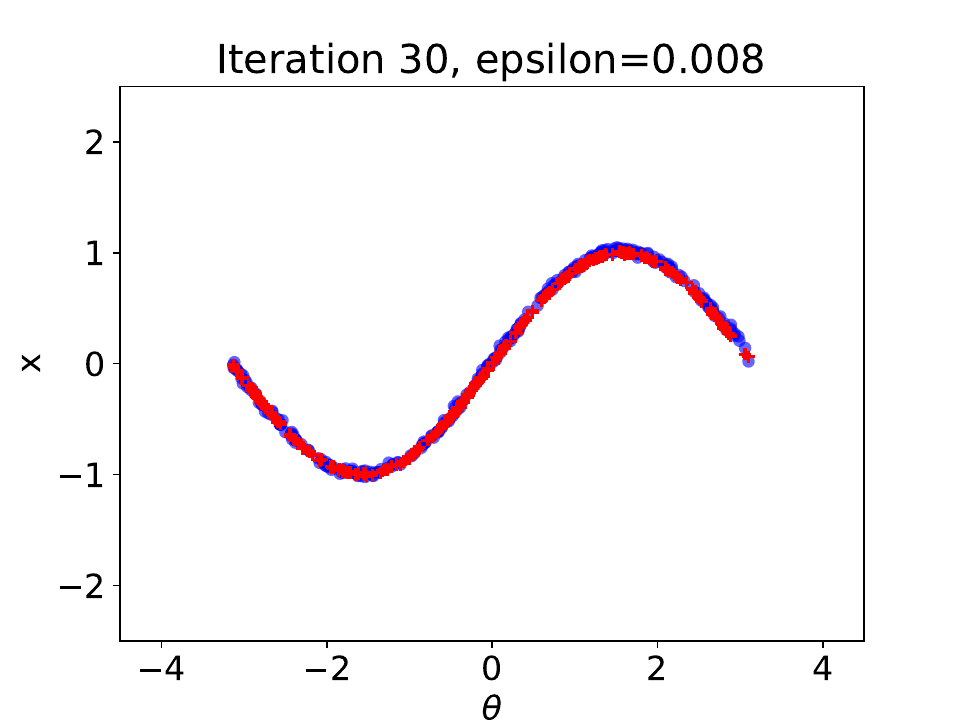}
\end{center}
\caption{Sinusoidal example output.
In each frame, dots are 300 samples from the current density $q$.
Crosses are 300 samples targeting $p_\epsilon(\theta)$.
These are based on the importance sampling stage of Algorithm \ref{alg:DIS} (steps 3--6)
using $q$ as a proposal, with importance resampling (see Section \ref{sec:improved grads}) used
to get unweighted samples.}
\label{fig:sin}
\end{figure}

Figure~\ref{fig:sin} shows our results.
The normalizing flow quickly adapts to meet the importance sampling results,
and after 30 iterations we reach $\epsilon=0.008$, taking roughly 0.1 minutes.
Visually, the samples lie close to the exact posterior support described above.
In some panels, $q$ has artefacts -- an unwanted ``tail'' at its right hand end --
which are removed by importance sampling.
This illustrates our recommendation (see start of Section \ref{sec:alg})
to use $q$ as an importance proposal, rather than as a posterior estimate.

\section{Example: M/G/1 Queue} \label{sec:MG1}

\subsection{Model}

Consider a M/G/1 queuing model of a single queue of customers.
Times between arrivals at the back of the queue are $\text{Exp}(\theta_1)$.
On reaching the front of the queue, a customer's service time is $U(\theta_2, \theta_3)$.
All these random variables are independent.
We consider a setting where only \emph{inter-departure times} are observed: times between departures from the queue.

We sample a synthetic dataset of $20$ observations from parameter values
$\theta_1 = 0.1, \theta_2 = 4, \theta_3 = 5$.
We attempt to infer these parameters under the prior
$\theta_1 \sim U(0,1/3), \theta_2 \sim U(0,10), \theta_3 - \theta_2 \sim U(0,10)$ (all independent).

\subsection{Existing Methods}

The M/G/1 model with inter-departure observations is a common benchmark for likelihood-free inference,
which can often provide fast approximate inference.
See \citet{Papamakarios:2018} for a comparison of methods for an M/G/1 example.
A potential disadvantage of existing likelihood-free methods is that they typically require using low dimensional summary statistics to be efficient,
and this can lose some information from the data.
As a likelihood-free inference baseline, we investigate a ABC-PMC algorithm.
This is a modification of existing ABC-PMC methods to use the unnormalised target \eqref{eq:LFItarget}.
See Section \ref{app:ABCbaseline} of the supplement for full details.

We also include results from a sophisticated MCMC scheme \citep{Shestopaloff:2014} for this model.
This provides near-exact samples from the posterior, which is a useful gold standard comparison for the approximate inference methods.
A limitation of this MCMC scheme is the difficulty in adapting it to other observation regimes
-- e.g.~observing the queue length at various times \citep{Pickands:1997} --
which would be relatively easy for likelihood-free methods.

\subsection{DIS Implementation}

We introduce $\vartheta$ and $x$,
vectors of length 3 and $2 \dim(y)$,
and take $\xi$ as the collection $(\vartheta, x)$.
Our simulator transforms these inputs to $\theta(\vartheta)$ and $y(\xi)$,
with the property that $\xi \sim \mathcal{N}(0,\mathrm{I})$ outputs a sample from the prior and the model.
See the supplement (Section \ref{app:MG1}) for details.
We pretrain $q$ to approximate this distribution.

\subsection{Results}

\begin{figure}[tbp]
\begin{center}
\includegraphics[width=0.49\textwidth]{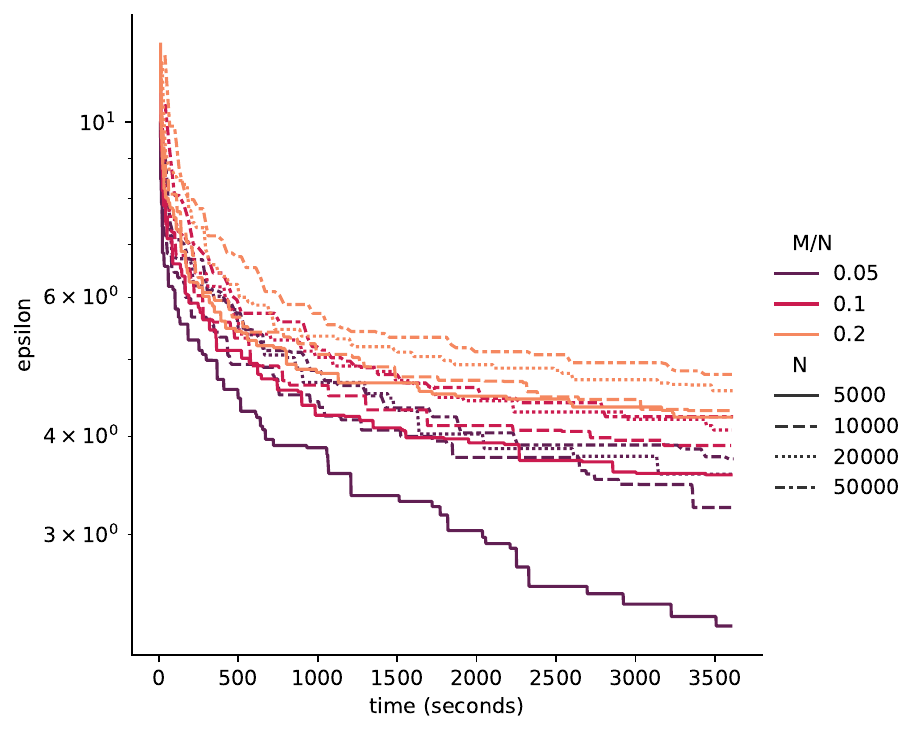}
\includegraphics[width=0.49\textwidth]{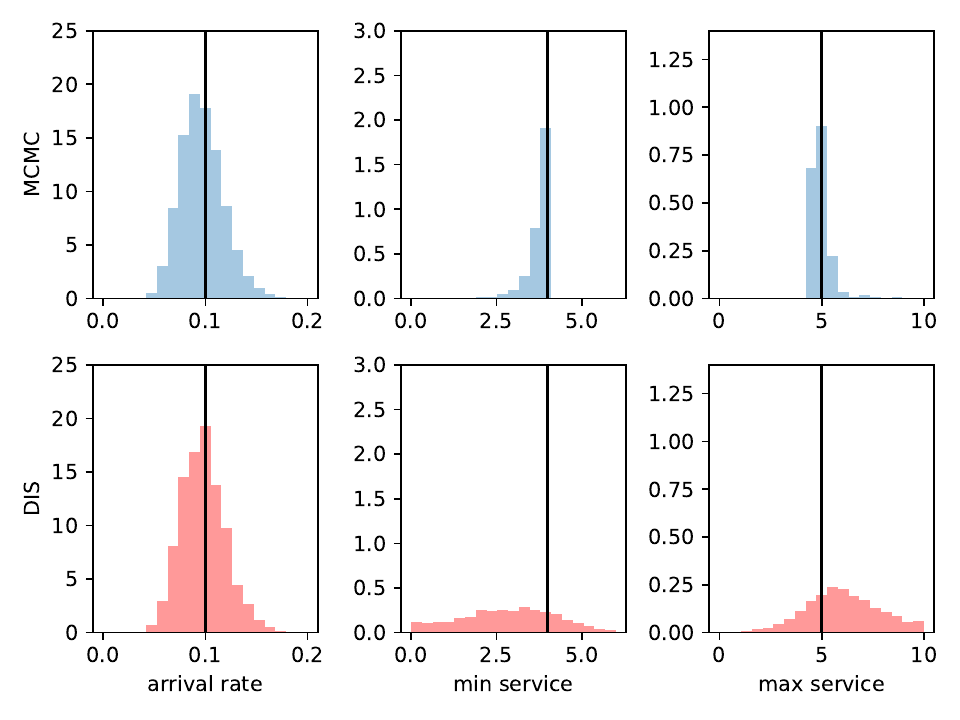}
\end{center}
\caption{
M/G/1 results.
Left:
The $\epsilon$ value reached by DIS on the M/G/1 example against computation time,
for various choices of $N$ (number of importance samples)
and $M/N$ (ratio of target effective sampling size to $N$).
Right:
Marginal posterior histograms for M/G/1 example.
The DIS output shown is for $\epsilon=2.30$,
which took 60 minutes to reach.
To produce DIS histograms we ran a final stage of importance sampling with $750,000$ samples,
and then used importance resampling (see Section \ref{sec:improved grads}) to get unweighted samples.
}
\label{fig:MG1}
\end{figure}

Figure~\ref{fig:MG1} (left) compares different choices of $N$ (number of importance samples) and $M$ (target ESS).
It shows that $\epsilon$ reduces more quickly for smaller $M/N$ and smaller $N$,
with $M/N$ having most influence.
Note that small $N$ also has the advantage of lower memory requirements.

This tuning study suggests taking $N$ and $M/N$ as small as possible,
while ensuring $M$ is at least a few hundred to avoid high variance in the importance sampling estimates.
We use this guidance here and in the next section's example.
In both these examples, the cost of a single simulator run is low.
For more expensive models efficient tuning choices may differ.

Figure~\ref{fig:MG1} (right)  is based on the DIS results with $N=5000$ and $M=250$,
following by a final importance sampling step with $750,000$ samples\footnotemark, which takes a further 7.2 minutes.
The results are a reasonably close match to near-exact MCMC output using the algorithm of \citet{Shestopaloff:2014}.
The DIS results for $\theta_2$ and $\theta_3$ are more diffuse than the MCMC results.
Also the $\theta_2$ DIS results lacks the sharp truncation of the right tail present in the MCMC results.
(Truncation occurs at the minimum observed inter-departure time of $4.01$,
as the minimum service time cannot be greater than this.)
\footnotetext{
The large number of samples is to meet a reviewer request to estimate the tails in Figure \ref{fig:MG1} accurately.
}

We also compare DIS to our ABC-PMC baseline, under a similar runtime.
For full details of methods and results see the supplement (Section \ref{app:ABCbaseline}).
Table~\ref{tab:MG1} summarizes the results.
ABC-PMC without summary statistics targets \eqref{eq:LFItarget}, just as DIS does,
but cannot reach as low an $\epsilon$ value in the same time,
so it produces inaccurate estimates for all parameters.
To improve ABC-PMC performance, we consider replacing the data with summaries:
quartiles of the inter-departure times, as used in \citet{Papamakarios:2018}.
This improves results for $\theta_1$ and $\theta_2$,
but still produces severe bias for $\theta_3$.

\begin{table}[tbp]
\begin{center}
\footnotesize
\begin{tabular}{c|ccc}
                     & $\theta_1$ (arrival rate) & $\theta_2$ (min service time) & $\theta_3$ (max service time) \\
  \hline
  MCMC               & 0.098 (0.060, 0.142)      & 3.72 (2.43, 4.00)             & 5.0  (4.4, 6.9)               \\
  ABC (no summaries) & 0.147 (0.070, 0.303)      & 6.72 (2.15, 9.47)             & 19.0 (12.4, 26.1)             \\
  ABC (summaries)    & 0.094 (0.056, 0.128)      & 3.82 (2.92, 4.26)             & 9.0  (7.9, 11.3)              \\
 
  ABC (no summaries) & 0.149 (0.066, 0.318)      & 6.20 (2.55, 9.63)             & 18.0 (10.6, 24.4)             \\
  ABC (summaries)    & 0.093 (0.056, 0.134)      & 3.80 (3.03, 4.33)             & 8.8  (7.6, 9.8)               \\
  DIS                & 0.098 (0.060, 0.145)      & 2.85 (0.20, 5.43)             & 6.4 (2.7, 11.1)
\end{tabular}
\end{center}
\caption{
  Estimated posterior means and 95\% quantile intervals
  for the M/G/1 example using various methods:
  MCMC (near-exact),
  ABC without summary statistics for $\epsilon=6.32$,
  ABC with quartile summaries for $\epsilon=0.16$,
  DIS for $\epsilon=2.30$.
  The $\epsilon$ values resulted from running ABC and DIS methods for roughly equal times.
}
\label{tab:MG1}
\end{table}

\section{Example: Susceptible-Infective Process on a Random Network} \label{sec:SI}
This section illustrates how DIS can be used to infer discrete variables.
The key idea is to represent them as transformations of continuous latent variables.
Such a mapping must be non-injective, and this is supported by the DIS algorithm without a need for any modifications.

We demonstrate this in the context of a spreading process model of an infectious disease on a random network,
specifically a susceptible-infective (SI) compartmental model on a Erd\H{o}s-R\'enyi network.
See \cite{Dutta:2018} for a discussion of similar models,
although not the exact one we consider.

\subsection{Model}
Our model represents a population of $m$ individuals as a network.
Each individual corresponds to a node.
The individuals are labelled $0,1,\ldots,m-1$.
An edge indicates contact between the corresponding individuals.
The network is assumed to be an instance of the Erd\H{o}s-R\'enyi random network \citep{Renyi:1959}, so each pair of nodes form an edge with probability $\theta_1$.

The spreading process of the infectious disease is modelled as a discrete-time process.
We consider a compartmental model where at any time each individual is in one of three states: susceptible, infective or immune.
Initially there is a single infective node, individual 0.
Susceptible neighbours of an infective individual at time $t$ are \emph{exposed}
and become infective at time $t+1$ with probability $\theta_2$.
Exposed individuals who are not infected instead become immune.
(All infection and edge formation events are assumed independent.)

Parameters $\theta_1$ and $\theta_2$ have independent $U(0,1)$ prior distributions.
We assume that the observations are a binary vector indicating infective status for each node/time combination.

\subsection{Likelihood-based Inference} \label{sec:SIexact}
The likelihood function of the model can be calculated by summing contributions from all possible networks, as detailed in the supplement (Section \ref{app:SI model}).
This is computationally feasible for small $m$,
enabling near-exact posterior inference.
Below, we use likelihood-based importance sampling as a benchmark for an example with $m=5$.
However, likelihood calculations become infeasibly expensive for larger $m$,
as the number of networks to sum is $2^{m(m-1)/2}$.
For instance below we also investigate an example with $m=10$,
giving $2^{45} \approx 3 \times 10^{13}$ total networks.

\subsection{DIS Implementation}
\paragraph{Latent Variables and Simulator}

Similarly to Section \ref{sec:MG1},
we introduce vectors of latent variables $\vartheta, x_{\text{edge}}, x_{\text{infect}}$.
Here $\dim(\vartheta) = 2, \dim(x_{\text{edge}})=\binom{m}{2}, \dim(x_{\text{infect}})=m$.
Now $\xi$ is the collection $(\vartheta, x_{\text{edge}}, x_{\text{infect}})$.

Our simulator transforms these inputs to $\theta(\vartheta)$ and $y(\xi)$.
As in Section \ref{sec:MG1}, the simulator has the property that
$\xi \sim \mathcal{N}(0,\mathrm{I})$ produces samples from the prior and model.
We pretrain $q$ to approximate this distribution.
Full details of the simulator transformation are in the supplement (Section \ref{app:SI model}).
In brief, $x_{\text{edge}}$ has an entry for each possible edge.
An edge occurs when the corresponding entry falls below a threshold controlled by $\vartheta_1$.
Similarly $x_{\text{infect}}$ has an entry for each individual.
When that individual is exposed,
then infection occurs if the corresponding entry falls below a threshold controlled by $\vartheta_2$.
(Only one random variable is needed per individual as they can only be exposed once.
Afterwards they are either infective or immune.)

To summarize, as well as continuous parameters $\theta$
we also wish to infer discrete variables: presence of edges and infection status.
As noted at the start of Section \ref{sec:SI}, the discrete variables are inferred
by representing them as a mapping of continuous latent variables $\xi$.

\subsection{Results}

\begin{figure}[tbp]
\centering
{\includegraphics[scale=0.5]{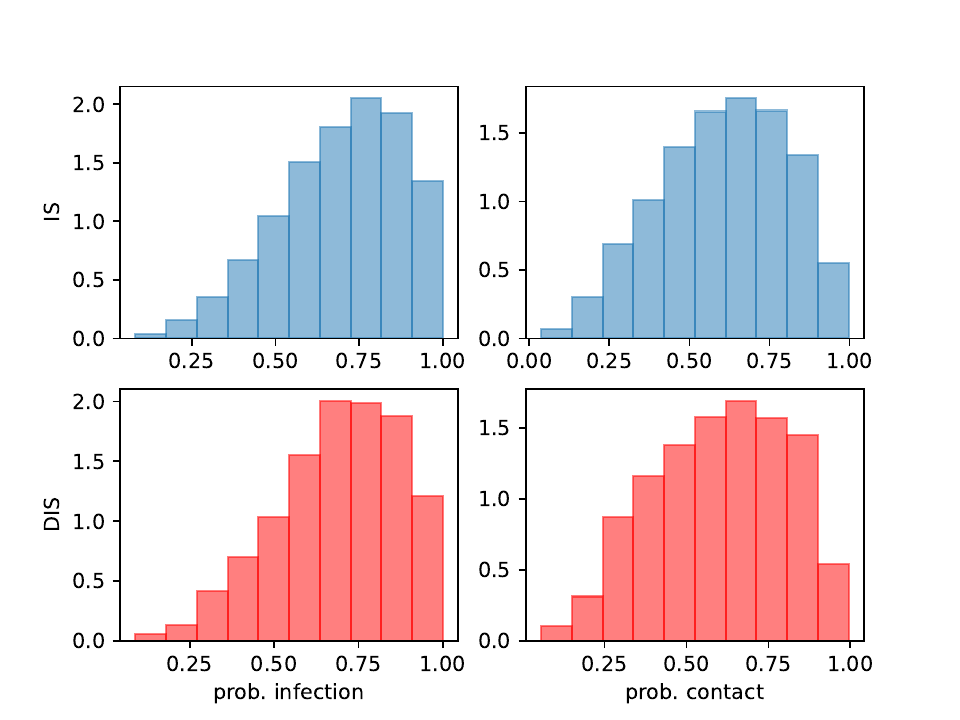}}
\caption{Output for SI example with $m=5$,
showing marginal posterior histograms for parameters $\theta_1$ and $\theta_2$
from IS and DIS output. 
To produce DIS histograms we ran a final stage of importance sampling with $100,000$ samples and then used importance resampling (see Section \ref{sec:improved grads}) to get unweighted samples.}
\label{fig:SI results1}
\end{figure}

We first consider a small network with $m=5$ nodes,
observed for $5$ time steps.
This has $\dim(\xi)=17$.
Following Section \ref{sec:MG1} we use $N=5000$ and $M=250$ in DIS.
As discussed in Section \ref{sec:LFItempering},
it is plausible here for DIS to reach $\epsilon=0$,
since the data are discrete.
Our experiment reaches $\epsilon=0$ after 1.4 minutes.
Then we perform importance sampling using this DIS proposal with $100,000$ samples, which takes a further 1.2 minutes.
Figure~\ref{fig:SI results1} shows posterior histograms for $\theta$ match those for likelihood-based importance sampling,
which produces $100,000$ samples in only 4 seconds.

\begin{figure}[tbp]
\raisebox{-0.45\height}{\includegraphics[width=0.49\textwidth]{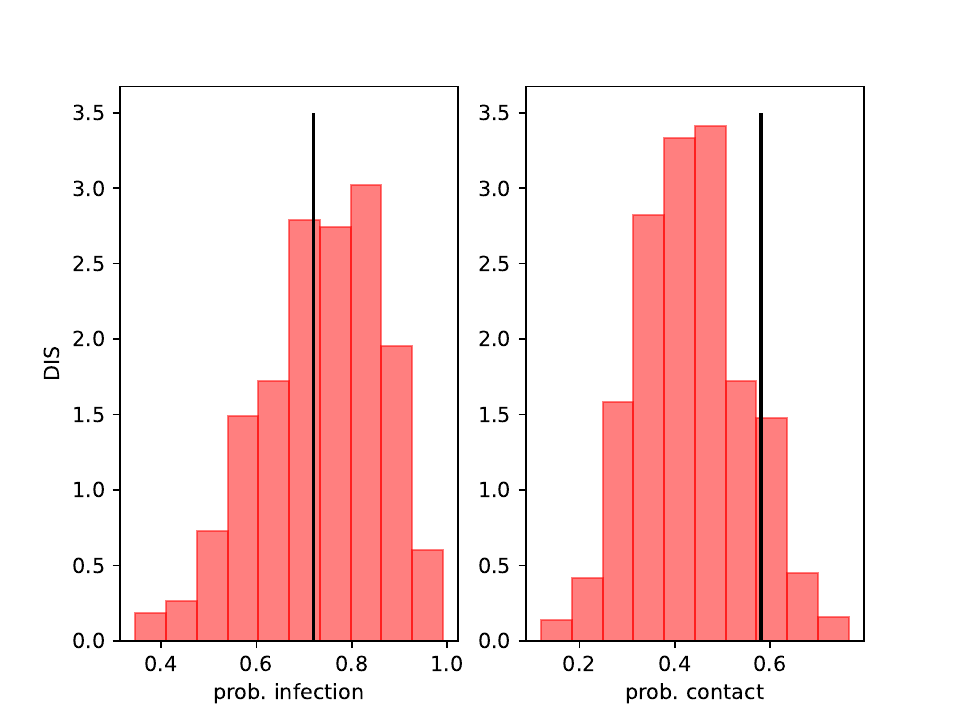}}
\raisebox{-0.45\height}
{\includegraphics[width=0.49\textwidth]{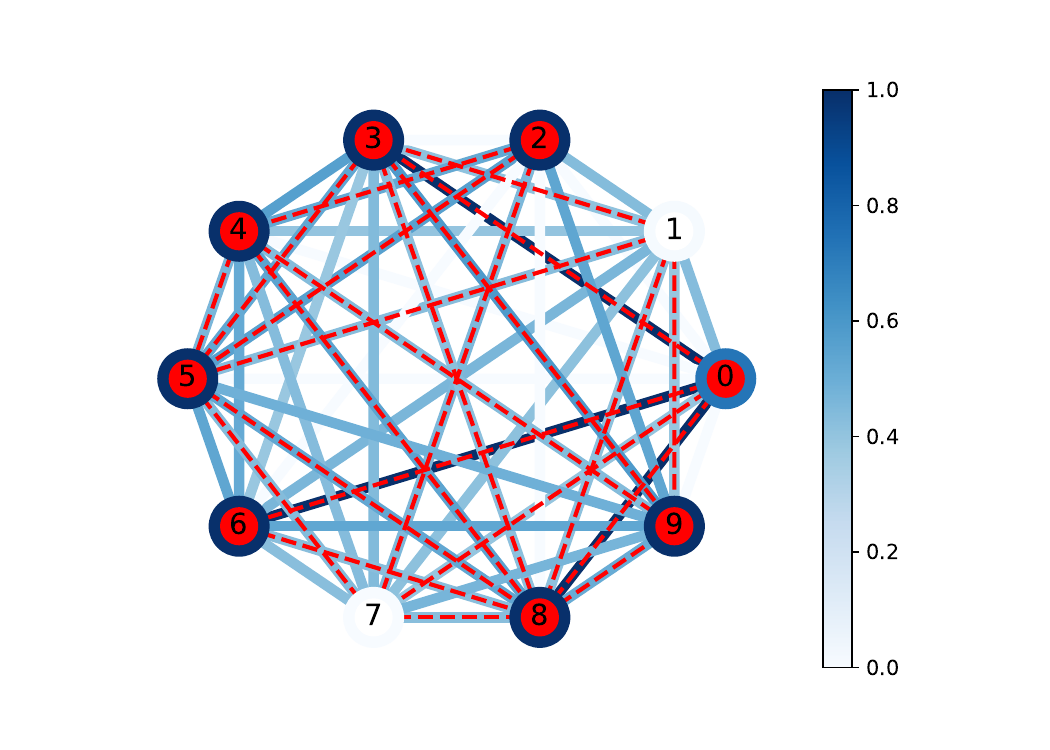}}
\caption{DIS output for SI example with $m=10$.
To produce the plots we ran a final stage of importance sampling with $20,000$ samples,
and then used importance resampling (see Section \ref{sec:improved grads}) to get unweighted samples.
Left: Marginal posterior histograms for parameters $\theta_1$ and $\theta_2$. Vertical lines show the true values.
Right: Posterior distribution of the network structure. Colours (in electronic version) /  shades (in printed version) indicate the posterior probability of the existence of an edge and that a node will be infected upon exposure. Nodes represented by shaded circles are infective at some point, and dashed edges are those existing in the true network.}
\label{fig:SI results2}
\end{figure}

Next we consider a larger network with $m=10$ nodes,
observed for $10$ time steps.
Now $\dim(\xi)=57$.
As described in Section \ref{sec:SIexact}, exact likelihood calculation is infeasible here.
However DIS with $N=5000$, $M=250$ can perform exact inference, reaching $\epsilon=0$ after 393 minutes.
Figure~\ref{fig:SI results2} (left) shows the resulting posterior histograms for $\theta$.
Then we perform importance sampling using this DIS proposal with $20,000$ samples, which takes a further 1.7 minutes.

DIS also outputs posterior distributions of the latent variables $x_{\text{edge}}$ and $x_{\text{infect}}$, controlling the network structure and whether individuals become immune on exposure.
These are summarized in Figure~\ref{fig:SI results2} (right).
In the observed data, only individuals 1 and 7 do not become infected.
The figure shows these individuals have a low posterior probability of becoming infected on exposure.
Individual 0 has a moderate probability of infection on exposure.
This is because they are already infective initially, so there is little information in the data on what would happen to them on exposure.
The remaining individuals have a high posterior probability of becoming infected.
In the data, individuals 3,6,8 are infected in the first time period.
Consequently the edges from these to individual 0 have the highest posterior probabilities,
as this is the only possible route of infection.
Edges representing plausible routes to individuals infected later have lower probabilities,
presumably as there are more possibilities for how this happens.
The lowest probability edges are those which would result in individuals being infected before this is observed to happen in the data e.g.~the edge $(0,2)$.

\section{Conclusion} \label{sec:conclusion}

We present \emph{distilled importance sampling} for likelihood-free inference,
and show its application in several examples.
An M/G/1 queuing example demonstrates that the method produces approximate results which are much more accurate than ABC,
without needing the use of summary statistics.
An epidemic model example shows that the method can also produce inference for discrete data.
Indeed it demonstrates that for discrete data, it is possible for DIS to target the exact posterior.
We also investigate tuning choices, and propose some default choices.
The supplement (Section \ref{sec:tuning summary}) has a summary of tuning recommendations,
and discussion of some alternative possibilities.

The remainder of the section discusses possible extensions of our methodology,
as well as limitations.

\paragraph{Likelihood-Based Inference}

This paper concentrates on likelihood-free inference.
DIS can also be applied to some likelihood-based inference problems.
However, exploratory work shows that here it perform less well than the competing methods
listed at the start of Section \ref{sec:related}.
See the supplement (Section \ref{sec:LBI}) for further discussion.


\paragraph{More Efficient Training}

Every iteration of Algorithm \ref{alg:DIS} generates new model simulations,
and uses them for a few iterations of training.
This results in a computationally feasible algorithm for our examples,
as the simulators are reasonably fast.
However for more expensive simulators, it would be desirable to use
simulations more efficiently in the training process.
For instance, training could continue with the same training data until convergence
-- although this risks overtraining.
Alternatively, training data from all previous iterations could be reused for training
-- although ensuring the correct target distribution may be difficult.
We plan to explore these approaches in future work.


\paragraph{Amortised Flows}

Throughout the paper, $q(\xi; \phi)$ is a generic normalizing flow producing the vector $\xi$.
We refer to this as a \emph{black-box} approach,
since it involves no knowledge of how $x$ is used by the simulator.
This is less practical for large $\dim(\xi)$,
as an increasingly large amount of simulations are required to learn such a high dimensional distribution.

An alternative is leveraging \emph{amortization} in the flow, using knowledge about the simulator.
Suppose a simulator requests a random variable.
An amortized flow would determine the distribution to sample from using knowledge of:
previously returned random variables (as for a black-box flow);
the current simulator state;
the line of code making the request.
The complexity of the flow now depends on the number of sampling statements in the code,
rather than the total number of samples made.
This can be a big reduction in complexity when the simulator makes many iterations over a few sample statements.
Similar ideas appear in \emph{inference networks} \citep{Le:2017}.

\paragraph{Overriding Random Number Generators}

For the examples in this paper, we wrote custom simulator code
which allowed $x$ values sampled from $q$ to be supplied as input.
For large simulator codebases, it would be ideal to work with the original code,
without needing to develop a custom version.
Following \citet{Baydin:2019},
a possible approach is to override the random number generator that the code uses
to instead provide $x$ values proposed by $q$.

\paragraph{Use with CDE}

As mentioned in Sections \ref{sec:intro} and \ref{sec:framework}, one approach to likelihood-free inference
is conditional density estimation (CDE).
Unlike DIS, this allows amortized inference:
after training it can be used for inference of many different observed datasets.
Future work could combine the methods:
use CDE to produce an initial approximate posterior estimate,
then refine further using DIS.
This makes use of complementary advantages of the methods:
CDE is amortized, while DIS relies less on summary statistics.

\paragraph{Acknowledgments}
The authors gratefully acknowledge: Alex Shestopaloff for providing MCMC code for the M/G/1 model;
Sammy Ragy for spotting several typos;
Andrew Golightly, Chris Williams and anonymous referees for helpful suggestions.
This work was supported by the EPSRC under grant EPSRC EP/V049127/1 (Dennis Prangle);
and Italian Ministry of University and Research (MIUR) under grant 58523\_DIPECC (Cecilia Viscardi).

{\centering \section*{Supplementary Material}}

\appendix

\renewcommand\thealgorithm{\Alph{algorithm}}

\section{Truncating Importance Weights} \label{app:truncation}

Importance sampling estimates can have large or infinite variance if
the proposal density $\lambda$ is a poor approximation to the target density $p$.
The practical manifestation of this is a small number of importance weights being very large relative to the others.
To reduce the variance, \citet{Ionides:2008} introduced \emph{truncated importance sampling}.
This replaces each importance weight $w_i$ with $\tilde{w}_i = \min(w_i, \omega)$
given some threshold $\omega$.
The truncated weights are then used in estimates \eqref{eq:ISunnorm} or \eqref{eq:ISnorm} (from the main paper).
Truncating in this way typically reduces variance, at the price of increasing bias.

\emph{Gradient clipping} \citep{Pascanu:2013} in stochastic gradient optimization operates by truncating large gradient estimates.
It is common practice to prevent occasional large gradient estimates from destabilizing optimization.
Truncating importance weights in Algorithm \ref{alg:DIS} (the DIS algorithm in the main paper)
has a similar effect of reducing the variability of gradient estimates.
A potential drawback of either method is that gradients lose the property of unbiasedness, which is theoretically required for convergence to an optimum of the objective.
\citet{Pascanu:2013} give heuristic arguments for good optimizer performance when using truncated gradients, and we make the following similar case for using truncated weights.
Firstly, even after truncation, the gradient is likely to point in a direction increasing the objective.
In our approach, it should still increase the $q$ density at $\xi^{(i)}$ values with large $w_i$ weights, which is desirable.
Secondly, we expect there is a region near the optimum for $\phi$ where truncation is extremely rare,
and therefore gradient estimates have very low bias once this region is reached.
Finally, we also observe good empirical behaviour in our examples, showing that optimization with truncated weights can find good importance sampling proposals.

Note that we could use gradient clipping directly in Algorithm \ref{alg:DIS}.
We prefer truncating importance weights as there is an automated way to choose the threshold $\omega$, as follows.
We select $\omega$ to reduce the maximum normalized importance weight,
$\max_{i} \frac{\tilde{w}_i}{\sum_{i=1}^N \tilde{w}_i}$,
to a prespecified value: throughout we use $0.1$.
The required $\omega$ can easily be calculated e.g.~by bisection.
(Occasionally no such $\omega$ exists i.e.~if most $w_i$s are zero.
In this case we set $\omega$ to the smallest positive $w_i$.)

\section{M/G/1 Model} \label{app:MG1}

This section describes the M/G/1 queuing model,
in particular how to simulate from it.

Recall that the parameters are $\theta_1, \theta_2, \theta_3$, with independent prior distributions $\theta_1 \sim U(0,1/3), \theta_2 \sim U(0,10), \theta_3 - \theta_2 \sim U(0,10)$.
We introduce a reparameterized version of our parameters: $\vartheta_1, \vartheta_2, \vartheta_3$ with independent $\mathcal{N}(0,1)$ priors.
Then we can take
$\theta_1 = \Phi(\vartheta_1)/3$,
$\theta_2 = 10 \Phi(\vartheta_2)$ and
$\theta_3 = \theta_2 + 10 \Phi(\vartheta_3)$,
where $\Phi$ is the $\mathcal{N}(0,1)$ cumulative distribution function.

The model involves independent latent variables $x_i \sim \mathcal{N}(0,1)$ for $1 \leq i \leq 2m$, where $m=\dim(y)$ is the number of observations.
These generate
\begin{align*}
a_i &= - \frac{1}{\theta_1} \log \Phi(x_i), & \text{(inter-arrival times)} \\
s_i &= \theta_2 + (\theta_3 - \theta_2) \Phi(x_{i+m}). & \text{(service times)}
\end{align*}
Inter-departure times can be calculated through the following recursion \citep{Lindley:1952}
\begin{align*}
d_i &= s_i + \max(A_i - D_{i-1}), & \text{(inter-departure times)}
\end{align*}
where $A_i = \sum_{j=1}^{i} a_j$ (arrival times)
and $D_i = \sum_{j=1}^{i} d_j$ (departure times).

\paragraph{Numerical Stability}

To avoid occasional numerical stability problems,
our implementation replaces the inter-arrival times equation above with
\[
a_i = \min \left\{ 10^6, - \frac{1}{\theta_1} \log [\Phi(x_i) + 10^{-20}] \right \}.
\]
We expect this to have negligible impact on the final results.

\section{ABC Analysis of M/G/1 Example} \label{app:ABCbaseline}

As a baseline comparison, we perform inference for the M/G/1 example using approximate Bayesian computation (ABC).
The algorithm and results are detailed here.

\subsection{Algorithm} \label{app:ABCalg}

We implement ABC using Algorithm \ref{alg:ABCPMC} below,
a modified version of the popular ABC-PMC approach \citep{Toni:2009, Sisson:2009, Beaumont:2009}.
These algorithms target
\[
\tilde{p}_\epsilon(\theta) =
\pi(\theta) \int \mathbbm{1}[||y(\xi)-y_0|| \leq \epsilon] \pi(x) dx,
\]
where $||\cdot||$ is the Euclidean norm,
and $\pi(x)$ is the density of the $x$ variables used in the simulator.
A novelty of Algorithm \ref{alg:ABCPMC} is that it instead targets
\[
\tilde{p}_\epsilon(\theta) =
\pi(\theta) \int \exp \left[ -\tfrac{1}{2 \epsilon^2} ||y(\xi)-y_0||^2 \right] \pi(x) dx.
\]
This is the $\theta$ marginal of the joint target used by DIS,
equation \eqref{eq:LFItarget} of the main paper.

Iteration $t$ of Algorithm \ref{alg:ABCPMC} produces
a weighted sample $(\theta^t_i, w^t_i)_{1 \leq i \leq N}$.
This targets $\tilde{p}_{\epsilon_t}(\theta)$ in the same way as importance sampling output.
Over the course of the algorithm, the value of $\epsilon_t$ is reduced,
and the number of simulated datasets required for an iteration tends to increase.

\begin{algorithm}
\caption{ABC-PMC}
\label{alg:ABCPMC}
\begin{algorithmic}[1]
\State initialize $\epsilon_1 = \infty$
\For{$t = 1,2,\ldots$}
\State Let $i=0$ (number of acceptances).
\While{$i < N$}
\State Sample $\theta^*$ from $\lambda_t(\theta)$,
defined in \eqref{eq:abcpmc_lambda}.
\If{$\pi(\theta^*) > 0$}
\State Sample $x^*$ from $\pi(x)$ and let $y^* = y(\theta^*, x^*)$.
\State Let $d^* = ||y^* - y_0||$.
\State Let $\alpha = \exp[-\tfrac{1}{2 \epsilon_t^2}{d^*}^2]$.
\State With probability $\alpha$ accept:
let $\theta^t_i=\theta^*$, $d^t_i = d^*$, $w^t_i = \pi(\theta^*)/\lambda_t(\theta^*)$
and increment $i$ by 1.
\EndIf
\EndWhile
\State Calculate $\epsilon_{t+1}$ (see below).
\EndFor
\end{algorithmic}
\end{algorithm}

Step 5 of Algorithm \ref{alg:ABCPMC} samples from the density $\lambda_t(\theta)$, where
\begin{subequations} \label{eq:abcpmc_lambda}
\begin{align}
\lambda_1(\theta) &= \pi(\theta) \\
\lambda_t(\theta) &= \frac{\sum_{i=1}^N w_i^{t-1} K_t(\theta | \theta_i^{t-1})}{\sum_{i=1}^N w_i^{t-1}} & \text{for } t>1.
\end{align}
\end{subequations}
In the first iteration $\lambda_t(\theta)$ is the prior.
After this, a kernel density estimate is used based on the previous weighted sample.
We follow \citet{Beaumont:2009} in using
\[
K_t(\theta | \theta^{t-1}) = \varphi(\theta^{t-1}, 2 \Sigma_{t-1}),
\]
where $\varphi$ is the density of a normal distribution and $\Sigma_{t-1}$ is the empirical variance matrix of $(\theta_i^{t-1})_{1 \leq i \leq N}$ calculated using weights $(w_i^{t-1})_{1 \leq i \leq N}$

To implement step 13 of the algorithm,
we select $\epsilon_{t+1}$ so that the acceptance probability of a typical member of the previous weighted sample is reduced by a prespecified factor $k$.
Specifically, we define $\tilde{d}$ as the median of the $d^t_i$ values,
and find $\epsilon_{t+1}$ by solving
\begin{align*}
\alpha(\tilde{d}, \epsilon_{t+1}) &= k \alpha(\tilde{d}, \epsilon_t),
\quad \text{where} \\
\alpha(d,\epsilon) &= \exp[-\tfrac{1}{2 \epsilon^2} d^2].
\end{align*}

\subsection{Results}

We run Algorithm \ref{alg:ABCPMC} on the M/G/1 example with $k=0.7$ and $N=250$.
Runtime is checked every time step 2 is reached,
and the algorithm terminates if this exceeds 70 minutes:
the time used by DIS for this example in the main paper
(including the final importance sampling stage).

Our implementation details give some modest advantages to ABC-PMC.
Firstly, under our stopping rule, the final iteration of ABC-PMC
typically continues several minutes beyond 70 minutes,
providing it with more computational resources than DIS.
Secondly, the ABC-PMC algorithm can perform more simulations per minute than DIS in this example.
This is because later ABC-PMC iterations require large numbers (millions) of simulations,
and our implementation was able to perform these in large parallel batches.

\begin{figure}[htb!]
\includegraphics[width=\textwidth]{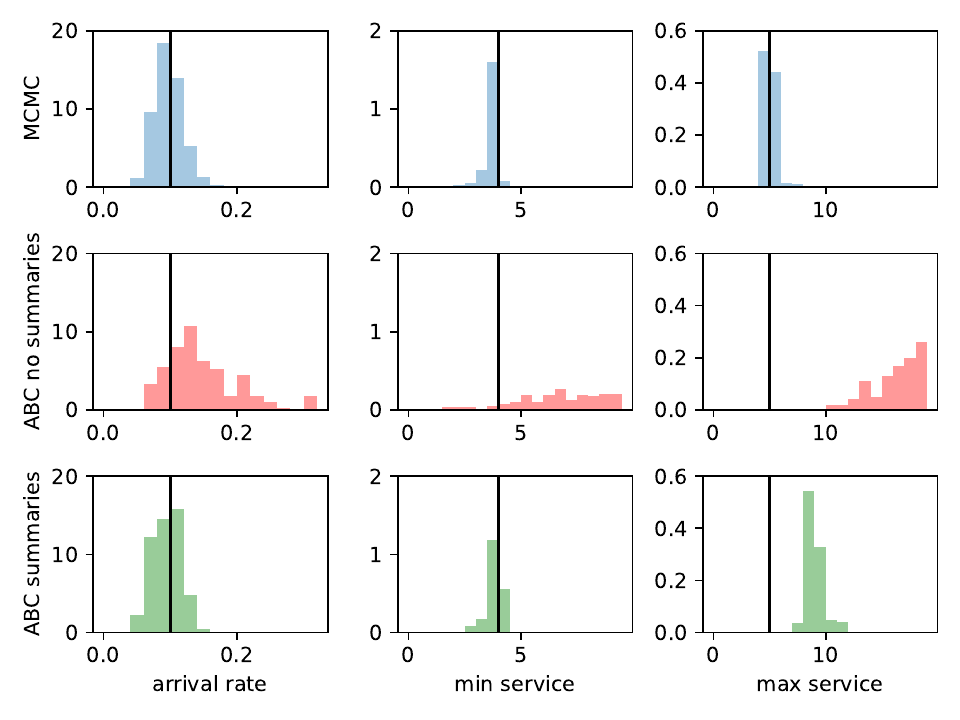}
\caption{
Marginal posterior histograms for M/G/1 example.
Top: MCMC output.
Middle: ABC output without summary statistics for $\epsilon=6.32$.
Bottom: ABC output with quartile summaries for $\epsilon=0.16$.
The ABC histograms are based on samples drawn from ABC-PMC output by importance resampling.
}
\label{fig:ABC posterior}
\end{figure}

Despite these advantages,
the final ABC-PMC $\epsilon$ value is $6.32$,
which is much larger than the final value for DIS, $\epsilon=2.30$.
Recall that the approximate posterior under $\epsilon$
is the exact posterior for data observed
with independent Gaussian errors of scale $\epsilon$ \citep{Wilkinson:2013}.
At $\epsilon=6.32$ this extra error is large compared to the data
(whose values range from 4 to 34),
hence it seems likely to produce significant approximation error.
This is reflected in very wide posterior marginals -- see Figure \ref{fig:ABC posterior}.
Further, the asymptotic cost per output sample of ABC is $O(\epsilon^{-\dim(y)})$
\citep{Prangle:2018},
suggesting that reaching $\epsilon=2.30$ using ABC requires a factor of
$(2.30/6.32)^{-20} \approx 6.0 \times 10^8$ more simulations, which is computationally infeasible.

To reduce the computational cost of ABC, the data are often replaced by low dimensional summary statistics.
We investigate using quartiles of the inter-departure times as summaries,
following \citet{Papamakarios:2016}.
This is done by adding a final step to the simulator which converts the raw data to quartiles,
and letting $y_0$ be the observed quartiles.
We run ABC-PMC with quartile statistics using the same tuning details as above.
The final $\epsilon$ value is now $0.16$.
Figure \ref{fig:ABC posterior} shows that the posterior marginals are much improved.
However the maximum service time marginal is still a poor approximation of the near-exact MCMC results,
suggesting that the summaries have lost information from the raw data.

\section{SI Process Example Further Details}\label{app:SI model}

In this section, we describe further details of our analysis of a SI process on a network.

\subsection{Simulator}

In this model, the parameters $\theta_1$ and $\theta_2$ represent the probability of contact and the probability of infection.
They are \emph{a priori} uniformly distributed on the unit interval.
Let $\vartheta_1$ and $\vartheta_2$ be two independent $\mathcal{N}(0,1)$ random variables.
We then take $\theta_1=\Phi(\vartheta_1)$ and $\theta_2=\Phi(\vartheta_2)$
where, as earlier,
$\Phi$ is the $\mathcal{N}(0,1)$ cumulative distribution function.

Recall that $m$ is the number of nodes in the network.
The latent variables $x_{\text{infect},1},\ldots,x_{\text{infect}, m}$ determine whether a node becomes infected or immune on its first exposure to infection.
The $i$-th node is infected at the first time period when it has a link to an infective node if $\Phi(x_{\text{infect}, i}) < \theta_2$
(or equivalently $x_{\text{infect}, i}< \vartheta_2$).
The latent variables $x_{\text{edge},1},\ldots,x_{\text{edge},m(m-1)/2}$ control network creation:
there are $m(m-1)/2$ possible edges and the $j$-th edge
(according to some enumeration of edges)
 is added when $\Phi(x_j)<\theta_1$
(or equivalently $x_j < \vartheta_1$).

Clearly when $\vartheta, x_{\text{edge}}, x_{\text{infect}}$ have a $\mathcal{N}(0,\mathrm{I})$ distribution,
then the simulator produces a sample from the prior and model.

\subsection{Likelihood-based Analysis } \label{app:SI ImpSamp}

Here we detail how the likelihood function for this model can be computed,
and outline why this is only computationally feasible for small $m$.

Our observations are the $I_{t,j}$ binary variables assuming value $1$ if the $j$-th node is infective at time $t$ and $0$ otherwise.
Here $j \in \{0,1,\ldots, m-1\}$ and $t \in \{0,1,2,\ldots,T-1\}$.
Define $I_t = (I_{t,0}, I_{t,1}, \ldots, I_{t,m-1})$.
We assume $I_0$ is fixed as $(1,0,0,\ldots,0)$.
So initially only individual 0 is infected.

Let us denote by $\mathcal{G}$ the set of adjacency matrices of size $m \times m$ for Erd\H{o}s-R\'enyi networks. The likelihood function is then
\begin{align}
\!\!\! L(\theta) &= \Pr( I_1,\ldots,I_{T-1}|\theta,I_0) \nonumber \\
&= \sum\limits_{G\in\mathcal{G}} \Pr( I_1,\ldots,I_{T-1},G|\theta,I_0) \nonumber \\
&= \sum\limits_{G\in\mathcal{G}} \! \Pr(G|\theta_1) \! \Pr( I_1,\ldots,I_{T-1}|G,\theta_2,I_0), \!
\label{eq:likelihood_SI}
\end{align}
where $\Pr(G|\theta_1)=\theta_1^{D}(1-\theta_1)^{\binom{m}{2}-D}$
with $D=\sum\limits_{i=1}^m\sum\limits_{j=i}^m G_{i,j}$, the degree of the network.

Let $A_1$ be the set of nodes neighbouring the seed node (individual 0).
For $t \geq 2$,
let $A_t=A(I_{t-1},I_{t-2})$ be the set $\{j:(\bar{I}_{t-2,t-1} G)_j >0\} \cap\{j: I_{t-1,j}=0\}$
where $\bar{I}_{t-2, t-1}= I_{t-1}\odot (1-I_{t-2})$.
(Recall $\odot$ represents elementwise multiplication.)
Thus $A_t$ is the set of nodes that could be infected at time $t$
i.e.~those susceptible at time $t-1$ and connected to a node that became infective at time $t-1$.
Then
\begin{align*}
& \Pr( I_1,\ldots,I_{T-1}|G,\theta_2,I_0)\\
&= \prod\limits_{t=2}^{T-1} \Pr(I_t|I_{t-1},I_{t-2},\theta_2,G)\Pr(I_1|I_0,\theta_2,G)\\
&=\prod\limits_{t=2}^{T-1} \prod\limits_{j=0}^{m-1}\Pr(I_{t,j}|I_{t-1},I_{t-2},\theta_2,G)\Pr(I_1|I_0,\theta_2,G)\\
&=\prod\limits_{t=1}^{T-1} \prod\limits_{j\in A_t} \theta_2^{I_{t,j}} (1-\theta_2)^{1-I_{t,j}}\prod\limits_{j\notin A_t}{ \mathbbm{1}\{I_{t-1,j}=I_{t,j}\}} .
\end{align*}
From \eqref{eq:likelihood_SI} it follows that with a small number of nodes the likelihood evaluation has an affordable computational cost.  However, the cost increases exponentially in $m$ because \eqref{eq:likelihood_SI} involves a summation over $\mathcal{G}$ having cardinality $2^{m(m-1)/2}$.

\section{Tuning Recommendations} \label{sec:tuning summary}

Here we summarize recommendations on tuning DIS which appear throughout the paper,
and add some further comments.

\subsection{Hyperparameters}

For the M/G/1 and SI network examples we recommend using:
\begin{itemize}
\item $N=5000$ (importance sampling sample size)
\item $M=250$ (target effective sample size)
\item $n=100$ (training batch size)
\item $B=\lceil M/n \rceil=3$ (number of batches)
\end{itemize}
The sinusoidal example is much simpler, and we expect a wide range of tuning choices to work well.
In the paper we use $N=4000, M=2000$ to allow easy visualization.

In general we expect optimizing these hyperparameters for particular tasks is likely to be useful.
In particular, this paper focuses on simulators which are computationally cheap to run.
For more expensive models, the optimal tuning choices could be qualitatively different.

\subsection{Normalizing Flow Architecture}

For all our examples we use an autoregressive neural spline flow for $q(\xi; \phi)$ with 5 bins for each variable, and bounding box $[-10,10]$.
The spline details are output by an autoregressive residual neural network \citep{Nash:2019}
with 3 blocks, 20 hidden features and ReLU activation.
We recommend this as a default choice of architecture.
However we expect more complex and higher dimensional targets to require larger architectures
e.g.~with more layers and hidden units.

An indication that a larger architecture is needed in the DIS algorithm is that $\epsilon_t$ stops decreasing
at some $\epsilon^*>0$.
This can occur when no density $q(\xi; \phi)$ from the current architecture is a sufficiently good approximation of $p_{\epsilon^*}$.
Therefore it becomes extremely unlikely for the target ESS value to be attained,
which is required to reduce $\epsilon_t$ further.

In exploratory work we also investigated the \emph{real NVP} (``Non-Volume Preserving'') normalizing flow \citep{Dinh:2016}.
We found this flow could be much faster to use than a spline flow
(i.e.~more iterations of Algorithm \ref{alg:DIS} can be run per unit time).
However it sometimes produced numerical errors in the M/G/1 example,
and struggled to approximate the target for low $\epsilon$ in the SI network example.
We speculate that the latter issue is because this posterior requires a complex dependency structure
-- e.g.~the presence of some contact edges rule out other edges --
that is easier to capture using an autoregressive flow.
Due to these issues, we recommend spline flows as a default choice.

\subsection{Selecting $\epsilon$}

The method outlined in the main paper to select $\epsilon$ is free of tuning parameters.
However, alternative methods to select $\epsilon$ could be used,
for instance by using variations on the standard effective sample size
(see e.g.~\citealp{Elvira:2022}).

\section{Likelihood-based inference} \label{sec:LBI}

Here we expand on comments from the main paper's conclusion, Section \ref{sec:conclusion},
on applying DIS to likelihood-based inference.
This is possible when a full data likelihood $L(\theta, x)$ is available given some latent variables $x$,
but integration to $L(\theta)$ is infeasibly computationally expensive.
In this case the unnormalised DIS target $\tilde{p}_\epsilon$ would be a tempered form of $\pi(\xi) L(\theta, x)$,
for instance $\tilde{p}_\epsilon(\xi) = \pi(\xi) L(\theta, x)^{1-\epsilon}$
(taking $\epsilon_0=1$).

Earlier drafts of this paper included such applications to likelihood-based inference.
A reviewer suggested investigating how useful tempering was.
Exploratory work found it was often of little use:
simply initialising $\epsilon_0=0$ in our algorithm produced good results.
The resulting algorithm has little novelty over other work listed at the start of Section \ref{sec:related}.
Therefore we have concentrated on LFI applications,
where initialising $\epsilon_0=0$ is generally not feasible
as it would result in all importance sampling weights being zero
(almost surely for continuous data, or with high probability for high dimensional discrete data).

Our exploratory findings about the poor performance of tempering for
likelihood-based applications of DIS may not apply to other related methods.
For instance, a reviewer commented that there is some similarity between DIS for likelihood-based inference
and recent work of \cite{Surjanovic:2022}.
This also uses a tempering scheme and estimates densities
by minimising the inclusive Kullback-Leibler divergence
(equation \eqref{eq:incKL} from the main paper).
Their work attains good empirical results.
However its goal is to create an efficient parallel tempering MCMC scheme,
rather than an importance sampling proposal as in DIS.

\bibliography{distilledIS}

\begin{thebibliography}{}

\bibitem[Agapiou et~al., 2017]{Agapiou:2017}
Agapiou, S., Papaspiliopoulos, O., Sanz-Alonso, D., and Stuart, A.~M. (2017).
\newblock Importance sampling: Intrinsic dimension and computational cost.
\newblock {\em Statistical Science}, 32(3):405--431.

\bibitem[Arbel et~al., 2021]{Arbel:2021}
Arbel, M., Matthews, A., and Doucet, A. (2021).
\newblock Annealed flow transport {Monte Carlo}.
\newblock In {\em International Conference on Machine Learning}, pages
  318--330.

\bibitem[Baydin et~al., 2018]{Baydin:2018}
Baydin, A.~G., Pearlmutter, B.~A., Radul, A.~A., and Siskind, J.~M. (2018).
\newblock Automatic differentiation in machine learning: a survey.
\newblock {\em Journal of Marchine Learning Research}, 18:1--43.

\bibitem[Baydin et~al., 2019]{Baydin:2019}
Baydin, A.~G., Shao, L., Bhimji, W., Heinrich, L., Naderiparizi, S., Munk, A.,
  Liu, J., Gram-Hansen, B., Louppe, G., Meadows, L., Torr, P., Lee, V.,
  Cranmer, K., Prabhat, and Wood, F. (2019).
\newblock Efficient probabilistic inference in the quest for physics beyond the
  standard model.
\newblock In {\em Advances in Neural Information Processing Systems},
  volume~32.

\bibitem[Beaumont et~al., 2009]{Beaumont:2009}
Beaumont, M.~A., Cornuet, J.-M., Marin, J.-M., and Robert, C.~P. (2009).
\newblock Adaptive approximate {B}ayesian computation.
\newblock {\em Biometrika}, 96(4):2025--2035.

\bibitem[Bornschein and Bengio, 2014]{Bornschein:2014}
Bornschein, J. and Bengio, Y. (2014).
\newblock Reweighted wake-sleep.
\newblock {\em arXiv preprint arXiv:1406.2751}.

\bibitem[Chatterjee and Diaconis, 2018]{Chatterjee:2018}
Chatterjee, S. and Diaconis, P. (2018).
\newblock The sample size required in importance sampling.
\newblock {\em The Annals of Applied Probability}, 28(2):1099--1135.

\bibitem[Cornuet et~al., 2012]{Cornuet:2012}
Cornuet, J.-M., Marin, J.-M., Mira, A., and Robert, C.~P. (2012).
\newblock Adaptive multiple importance sampling.
\newblock {\em Scandinavian Journal of Statistics}, 39(4):798--812.

\bibitem[Cotter et~al., 2020]{Cotter:2019}
Cotter, S.~L., Kevrekidis, I.~G., and Russell, P. (2020).
\newblock Transport map accelerated adaptive importance sampling, and
  application to inverse problems arising from multiscale stochastic reaction
  networks.
\newblock {\em SIAM/ASA Journal on Uncertainty Quantification},
  8(4):1383--1413.

\bibitem[Del~Moral et~al., 2012]{DelMoral:2012}
Del~Moral, P., Doucet, A., and Jasra, A. (2012).
\newblock An adaptive sequential {Monte Carlo} method for approximate
  {Bayesian} computation.
\newblock {\em Statistics and Computing}, 22(5):1009--1020.

\bibitem[Dieng et~al., 2017]{Dieng:2017}
Dieng, A.~B., Tran, D., Ranganath, R., Paisley, J., and Blei, D. (2017).
\newblock Variational inference via $\chi$ upper bound minimization.
\newblock In {\em Advances in Neural Information Processing Systems},
  volume~30.

\bibitem[Dinh et~al., 2016]{Dinh:2016}
Dinh, L., Sohl-Dickstein, J., and Bengio, S. (2016).
\newblock Density estimation using real {NVP}.
\newblock {\em arXiv preprint arXiv:1605.08803}.

\bibitem[Duan, 2021]{Duan:2019}
Duan, L.~L. (2021).
\newblock Transport {Monte Carlo}: high-accuracy posterior approximation via
  random transport.
\newblock {\em Journal of the American Statistical Association (online
  preview)}.

\bibitem[Durkan et~al., 2019]{Durkan:2019}
Durkan, C., Bekasov, A., Murray, I., and Papamakarios, G. (2019).
\newblock Neural spline flows.
\newblock In {\em Advances in Neural Information Processing Systems},
  volume~32.

\bibitem[Dutta et~al., 2018]{Dutta:2018}
Dutta, R., Mira, A., and Onnela, J.-P. (2018).
\newblock Bayesian inference of spreading processes on networks.
\newblock {\em Proceedings of the Royal Society A: Mathematical, Physical and
  Engineering Sciences}, 474(2215):20180129.

\bibitem[Elvira et~al., 2022]{Elvira:2022}
Elvira, V., Martino, L., and Robert, C.~P. (2022).
\newblock Rethinking the effective sample size.
\newblock {\em International Statistical Review}.

\bibitem[Erd{\H{o}}s and R{\'e}nyi, 1959]{Renyi:1959}
Erd{\H{o}}s, P. and R{\'e}nyi, A. (1959).
\newblock On random graphs.
\newblock {\em Publicationes Mathematicate}, 6:290--297.

\bibitem[Foerster et~al., 2018]{Foerster:2018}
Foerster, J., Farquhar, G., Al-Shedivat, M., Rockt{\"a}schel, T., Xing, E., and
  Whiteson, S. (2018).
\newblock Dice: The infinitely differentiable {Monte Carlo} estimator.
\newblock In {\em International Conference on Machine Learning}, pages
  1529--1538.

\bibitem[Germain et~al., 2015]{Germain:2015}
Germain, M., Gregor, K., Murray, I., and Larochelle, H. (2015).
\newblock {MADE}: Masked autoencoder for distribution estimation.
\newblock In {\em International Conference on Machine Learning}, pages
  881--889.

\bibitem[Graham and Storkey, 2017]{Graham:2017}
Graham, M.~M. and Storkey, A.~J. (2017).
\newblock Asymptotically exact inference in differentiable generative models.
\newblock {\em Electronic Journal of Statistics}, 11(2):5105--5164.

\bibitem[Grazian and Fan, 2019]{Grazian:2019}
Grazian, C. and Fan, Y. (2019).
\newblock A review of approximate {B}ayesian computation methods via density
  estimation: Inference for simulator-models.
\newblock {\em Wiley Interdisciplinary Reviews: Computational Statistics},
  12(4):e1486.

\bibitem[Huggins et~al., 2020]{Huggins:2020}
Huggins, J.~H., Kasprzak, M., Campbell, T., and Broderick, T. (2020).
\newblock Practical posterior error bounds from variational objectives.
\newblock In {\em Artificial Intelligence and Statistics}, pages 1792--1802.

\bibitem[Ikonomov and Gutmann, 2020]{Ikonomov:2020}
Ikonomov, B. and Gutmann, M.~U. (2020).
\newblock Robust optimisation {Monte Carlo}.
\newblock In {\em International Conference on Artificial Intelligence and
  Statistics}, pages 2819--2829.

\bibitem[Ionides, 2008]{Ionides:2008}
Ionides, E.~L. (2008).
\newblock Truncated importance sampling.
\newblock {\em Journal of Computational and Graphical Statistics},
  17(2):295--311.

\bibitem[Jerfel et~al., 2021]{Jerfel:2021}
Jerfel, G., Wang, S., Fannjiang, C., Heller, K.~A., Ma, Y., and Jordan, M.~I.
  (2021).
\newblock Variational refinement for importance sampling using the forward
  {Kullback-Leibler} divergence.
\newblock In {\em Uncertainty in Artificial Intelligence}, pages 1819--1829.

\bibitem[Kingma and Ba, 2015]{Kingma:2014}
Kingma, D.~P. and Ba, J. (2015).
\newblock Adam: A method for stochastic optimization.
\newblock In {\em International Conference on Learning Representations}.

\bibitem[Larra{\~n}aga and Lozano, 2002]{Larranaga:2002}
Larra{\~n}aga, P. and Lozano, J.~A. (2002).
\newblock {\em Estimation of distribution algorithms: A new tool for
  evolutionary computation}.
\newblock Springer.

\bibitem[Le et~al., 2017]{Le:2017}
Le, T.~A., Baydin, A.~G., and Wood, F. (2017).
\newblock Inference compilation and universal probabilistic programming.
\newblock In {\em Artificial Intelligence and Statistics}, pages 1338--1348.

\bibitem[Li et~al., 2017]{Li:2017}
Li, Y., Turner, R.~E., and Liu, Q. (2017).
\newblock Approximate inference with amortised {MCMC}.
\newblock {\em arXiv preprint arXiv:1702.08343}.

\bibitem[Lindley, 1952]{Lindley:1952}
Lindley, D.~V. (1952).
\newblock The theory of queues with a single server.
\newblock {\em Mathematical Proceedings of the Cambridge Philosophical
  Society}, 48(2):277--289.

\bibitem[Liu, 1996]{Liu:1996}
Liu, J.~S. (1996).
\newblock Metropolized independent sampling with comparisons to rejection
  sampling and importance sampling.
\newblock {\em Statistics and Computing}, 6(2):113--119.

\bibitem[MacKay, 2003]{Mackay:2003}
MacKay, D. J.~C. (2003).
\newblock {\em Information theory, inference and learning algorithms}.
\newblock Cambridge University Press.

\bibitem[Marin et~al., 2012]{Marin:2012}
Marin, J.-M., Pudlo, P., Robert, C.~P., and Ryder, R.~J. (2012).
\newblock Approximate {Bayesian} computational methods.
\newblock {\em Statistics and Computing}, 22(6):1167--1180.

\bibitem[Meeds and Welling, 2015]{Meeds:2015}
Meeds, T. and Welling, M. (2015).
\newblock Optimization {Monte Carlo}: Efficient and embarrassingly parallel
  likelihood-free inference.
\newblock {\em Advances in Neural Information Processing Systems}, 28.

\bibitem[Mohamed et~al., 2020]{Mohamed:2019}
Mohamed, S., Rosca, M., Figurnov, M., and Mnih, A. (2020).
\newblock {Monte Carlo} gradient estimation in machine learning.
\newblock {\em Journal of Machine Learning Research}, 21(132):1--62.

\bibitem[M{\"u}ller et~al., 2019]{Muller:2019}
M{\"u}ller, T., Mcwilliams, B., Rousselle, F., Gross, M., and Nov{\'a}k, J.
  (2019).
\newblock Neural importance sampling.
\newblock {\em ACM Transactions on Graphics (TOG)}, 38(5):1--19.

\bibitem[Naesseth et~al., 2021]{Naesseth:2020}
Naesseth, C.~A., Lindsten, F., and Blei, D. (2021).
\newblock Markovian score climbing: Variational inference with {$KL(p||q)$}.
\newblock In {\em Advances in Neural Information Processing Systems},
  volume~33, pages 15499--15510.

\bibitem[Nash and Durkan, 2019]{Nash:2019}
Nash, C. and Durkan, C. (2019).
\newblock Autoregressive energy machines.
\newblock In {\em International Conference on Machine Learning}, pages
  1735--1744.

\bibitem[Papamakarios and Murray, 2016]{Papamakarios:2016}
Papamakarios, G. and Murray, I. (2016).
\newblock Fast $\varepsilon$-free inference of simulation models with
  {B}ayesian conditional density estimation.
\newblock In {\em Advances in Neural Information Processing Systems},
  volume~29.

\bibitem[Papamakarios et~al., 2021]{Papamakarios:2019}
Papamakarios, G., Nalisnick, E., Rezende, D.~J., Mohamed, S., and
  Lakshminarayanan, B. (2021).
\newblock Normalizing flows for probabilistic modeling and inference.
\newblock {\em Journal of Machine Learning Research}, 22(57):1--64.

\bibitem[Papamakarios et~al., 2019]{Papamakarios:2018}
Papamakarios, G., Sterratt, D., and Murray, I. (2019).
\newblock Sequential neural likelihood: Fast likelihood-free inference with
  autoregressive flows.
\newblock In {\em Artificial Intelligence and Statistics}, pages 837--848.

\bibitem[Pascanu et~al., 2013]{Pascanu:2013}
Pascanu, R., Mikolov, T., and Bengio, Y. (2013).
\newblock On the difficulty of training recurrent neural networks.
\newblock In {\em International Conference on Machine Learning}, pages
  1310--1318.

\bibitem[Paszke et~al., 2019]{Paszke:2019}
Paszke, A., Gross, S., Massa, F., Lerer, A., Bradbury, J., Chanan, G., Killeen,
  T., Lin, Z., Gimelshein, N., Antiga, L., et~al. (2019).
\newblock Pytorch: An imperative style, high-performance deep learning library.
\newblock In {\em Advances in Neural Information Processing Systems},
  volume~32.

\bibitem[Pickands and Stine, 1997]{Pickands:1997}
Pickands, III, J. and Stine, R.~A. (1997).
\newblock Estimation for an {M/G/$\infty$} queue with incomplete information.
\newblock {\em Biometrika}, 84(2):295--308.

\bibitem[Prangle et~al., 2018]{Prangle:2018}
Prangle, D., Everitt, R.~G., and Kypraios, T. (2018).
\newblock A rare event approach to high-dimensional approximate {Bayesian}
  computation.
\newblock {\em Statistics and Computing}, 28(4):819--834.

\bibitem[Robert and Casella, 2013]{Robert:2013}
Robert, C.~P. and Casella, G. (2013).
\newblock {\em {M}onte {C}arlo statistical methods}.
\newblock Springer.

\bibitem[Rubinstein, 1999]{Rubinstein:1999}
Rubinstein, R. (1999).
\newblock The cross-entropy method for combinatorial and continuous
  optimization.
\newblock {\em Methodology and computing in applied probability},
  1(2):127--190.

\bibitem[Rubinstein and Kroese, 2016]{Rubinstein:2016}
Rubinstein, R.~Y. and Kroese, D.~P. (2016).
\newblock {\em Simulation and the {Monte Carlo} method}.
\newblock John Wiley \& Sons.

\bibitem[Ruder, 2016]{Ruder:2016}
Ruder, S. (2016).
\newblock An overview of gradient descent optimization algorithms.
\newblock {\em arXiv preprint arXiv:1609.04747}.

\bibitem[Shestopaloff and Neal, 2014]{Shestopaloff:2014}
Shestopaloff, A.~Y. and Neal, R.~M. (2014).
\newblock On {Bayesian} inference for the {M/G/1} queue with efficient {MCMC}
  sampling.
\newblock {\em arXiv preprint arXiv:1401.5548}.

\bibitem[Sisson et~al., 2009]{Sisson:2009}
Sisson, S.~A., Fan, Y., and Tanaka, M.~M. (2009).
\newblock Correction: {Sequential Monte Carlo} without likelihoods.
\newblock {\em Proceedings of the National Academy of Sciences},
  106(39):16889--16890.

\bibitem[Smith and Gelfand, 1992]{Smith:1992}
Smith, A. F.~M. and Gelfand, A.~E. (1992).
\newblock Bayesian statistics without tears: a sampling--resampling
  perspective.
\newblock {\em The American Statistician}, 46(2):84--88.

\bibitem[Surjanovic et~al., 2022]{Surjanovic:2022}
Surjanovic, N., Syed, S., Bouchard-C{\^o}t{\'e}, A., and Campbell, T. (2022).
\newblock Parallel tempering with a variational reference.
\newblock {\em arXiv preprint arXiv:2206.00080}.

\bibitem[Toni et~al., 2009]{Toni:2009}
Toni, T., Welch, D., Strelkowa, N., Ipsen, A., and Stumpf, M. (2009).
\newblock Approximate {B}ayesian computation scheme for parameter inference and
  model selection in dynamical systems.
\newblock {\em Journal of The Royal Society Interface}, 6(31):187--202.

\bibitem[Wilkinson, 2013]{Wilkinson:2013}
Wilkinson, R.~D. (2013).
\newblock Approximate {Bayesian} computation ({ABC}) gives exact results under
  the assumption of model error.
\newblock {\em Statistical applications in genetics and molecular biology},
  12(2):129--141.

\bibitem[Yao et~al., 2018]{Yao:2018}
Yao, Y., Vehtari, A., Simpson, D., and Gelman, A. (2018).
\newblock Yes, but did it work?: Evaluating variational inference.
\newblock In {\em International Conference on Machine Learning}, pages
  5581--5590.

\end{thebibliography}

\end{document}